\documentstyle[12pt,bezier]{article}
\begin{document}
\newcommand{\de}{\delta}\newcommand{\ga}{\gamma}\newcommand{\e}{\epsilon}
\newcommand{\th}{\theta}\newcommand{\ot}{\otimes}
\newcommand{\ba}{\begin{array}} \newcommand{\ea}{\end{array}}
\newcommand{\beq}{\begin{equation}}\newcommand{\eeq}{\end{equation}}
\newcommand{\tmod}{{\cal T}}\newcommand{\amod}{{\cal A}}
\newcommand{\bemod}{{\cal B}}\newcommand{\cmod}{{\cal C}}
\newcommand{\dmod}{{\cal D}}\newcommand{\hmod}{{\cal H}}
\newcommand{\s}{\scriptstyle}\newcommand{\tr}{{\rm tr}}
\newcommand{\einsop}{{\bf 1}}
\title{Quantum Group Invariant Integrable n-State Vertex Models with
Periodic Boundary Conditions}
\author{M. Karowski\thanks{e-mail: karowski@vaxneu.dnet.fu-berlin.de}
\\A. Zapletal\thanks{e-mail: zapletal@risc4.physik.fu-berlin.de}}
\date{\small\it Institut f\"{u}r Theoretische Physik\\
Freie Universit\"{a}t Berlin\\Germany}
\maketitle
\begin{abstract}
An $U_q(sl(n))$ invariant transfer matrix with periodic boundary conditions
is analysed by means of the algebraic nested Bethe ansatz for the case of
$q$ being a root of unity. The transfer matrix corresponds to a
2-dimensional vertex model on a torus with topological interaction w.r.t. the
3-dimensional interior of the torus. By means of finite size analysis we
find the central charge of the corresponding Virasoro algebra as
$c=(n-1) \left[1-n(n+1)/(r(r-1))\right] $.
\end{abstract}
\section{Introduction}
Since Bethe's pioneering work \cite{bethe} on the isotropic XXX-Heisenberg
model more than sixty years ago, the Bethe ansatz method has become one
of the most important
tools in analysing one dimensional integrable quantum chains or
equivalently 2-dimensional statistical models. Moreover,
it turned out that there is a deep
connection between the Bethe ansatz and the
underlying symmetry group of the model.
This has been stressed
by Faddeev and Takhadzhyan \cite{fad} investigating the Heisenberg model.
For general simple Lie groups this algebraic structure has been summarized in
 ref. \cite{wieg}.

For the case of the anisotropic XXZ-Heisenberg model
and generalizations of it the
underlying Yang-Baxter algebras, which guarantee the integrability of the
system, are related to new mathematical structures.
Drinfeld \cite{drinfeld} and Jimbo \cite{jimbo2}
have formulated these new structures as quantum groups.
Therefrom the question arises whether quantum groups should serve as a
generalization of symmetry concepts in physics.

However, deforming the $SU(2)$-invariant XXX-Heisenberg model with
periodic boundary conditions in the traditional \cite{yang} way
one obtains an XXZ-Hamiltonian which is not $U_q(sl(2))$-invariant
\cite{pasquier}, \cite{karowskieth}.
The reason for this puzzle is the non-cocommutativity of quantum
groups. This means, roughly speaking, for tensor products one has
to distinguish between left and right. So it is not obvious how to
identify the most left lattice point with most right one in order
to have periodic boundary conditions.

One possibility to obtain a quantum group invariant XXZ-Hamiltonian
is to consider open boundary conditions.
Such Hamiltonians have been
investigated by several authors (see e.g.
\cite{alcaraz}, \cite{pasquier}, \cite{ritten})\footnote{
Sometimes these models are formulated such that the open boundary
conditions are not quite obvious.}.
For open boundary conditions one has to apply Bethe ansatz techniques
introduced by Sklyanin \cite{skly} using Cherednik's
\cite{cher} 'reflection property'.
By this method the XXZ-Heisenberg model (see e.g.~\cite{devega2q}),
the $spl_q(2,1)$ invariant
supersymmetric t-J model \cite{forster} and the $U_q(sl(n))$ invariant
generalization of the XXZ-chain \cite{deveganq} have been solved for
open boundary conditions.
It turns out that these computations are quite involved, especially
for the nested Bethe ansatz case.

\begin{figure}[h]
\begin{center}
\setlength{\unitlength}{.5mm}
\begin{picture}(60.,70.)
%f1
\put(5.,60.25){\line(0,-1){45.}}
\put(50.,15.25){\line(0,1){45.}}
\bezier{150}(5.,15.25)(27.50,0.25)(50.,15.25)
\bezier{150}(5.,60.25)(27.50,45.25)(50.,60.25)
\bezier{150}(5.,60.25)(27.50,75.25)(50.,60.25)
\thicklines
\bezier{100}(5.,23.)(27.50,8.)(49.50,23.)
\bezier{100}(5.,37.50)(27.50,22.50)(49.50,37.50)
\bezier{100}(5.,53.)(27.50,38.)(49.50,53.)
\put(10.,57.20){\line(0,-1){45.}}
\put(20.,53.60){\line(0,-1){45.}}
\put(35.,53.60){\line(0,-1){45.}}
\put(45.,57.20){\line(0,-1){45.}}
\thinlines
\put(6.50,2.25){\makebox(6.50,6.){$2$}}
\put(16.75,1.25){\makebox(6.25,3.75){$1$}}
\put(32.50,1.75){\makebox(5.25,3.){$N$}}
\put(50.50,10.75){\makebox(2.,1.75){$\cdot$}}
\put(3.75,7.25){\makebox(2.,1.75){$\cdot$}}
\put(0.75,10.25){\makebox(2.,1.75){$\cdot$}}
\put(41.75,4.50){\makebox(2.,1.75){$\cdot$}}
\put(46.50,7.25){\makebox(2.,1.75){$\cdot$}}
\end{picture}
\end{center}
\caption{\label{f1}\it Square lattice on a cylinder considered as a
part of a torus}
\end{figure}
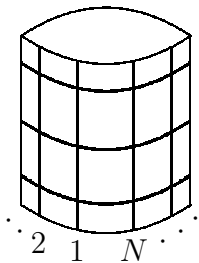

In the following we present a type of models with periodic
boundary conditions which are in addition quantum group invariant.
We  consider an n-state vertex model on an $N\times M$-square lattice
on a torus or cylinder as shown in Fig.~\ref{f1}.
As usual the partition function may be written as
%//{1.0}
\begin{equation}\label{1.0}
Z={\rm Tr}\ \tau^M,
\end{equation}
where the transfer matrix $\tau$ maps one cyclic chain $N,N-1,\dots,2,1$
to the next.
As is well known, equivalently to these types of model one may analyse
one dimensional quantum chain models.
We consider a cyclic chain $N,N-1,\dots,2,1,N,\dots$ where
a fundamental representation space
$V^{\Lambda_1}={\bf C}^n$ of $U_q(sl(n))$ is associated
to each lattice point $i$, ($i=N,\dots,1$).
As a generalization of the Heisenberg model with periodic boundary
conditions one can write an integrable $U_q(sl(n))$-symmetric Hamiltonian
%//{1.1}
\begin{equation}\label{1.1}
H=P^{\Lambda_2}_{NN-1}+\dots+P^{\Lambda_2}_{21}+P^{\Lambda_2}_{1N}
\end{equation}
acting in the tensor product space
%//{1.2}
\begin{equation}\label{1.2}
\Omega=V^{\Lambda_1}_N\otimes\cdots\otimes V^{\Lambda_1}_1.
\end{equation}
The $P^{\Lambda_2}_{ik}$ in eq.~(\ref{1.1}) project onto the
representation $\Lambda_2$ contained in the product
$\Lambda_1\otimes\Lambda_1$ associated to the lattice points
$i$ and $k$ according to
%//{1.3}
\begin{equation}\label{1.3}
\Lambda_1\otimes\Lambda_1=2\Lambda_1\oplus\Lambda_2 .
\end{equation}
For the case of $q=1$ ($SU(n)$-symmetry) these projectors
$P^{\Lambda_2}_{ik}$ are symmetric under
the exchange of $i$ and $k$. Therefore it is obvious how to define
$P^{\Lambda_2}_{1N}$. This changes for the quantum group case.
The q-projectors can be written in terms of the unit matrices
in $M_{n,n}(\bf C)$ for $i>k$ (lattice point $i$ on the left
of lattice point $k$) (see also \cite{ritten})
%
%//{1.4}
\begin{equation}\label{1.4}
P^{\Lambda_2}_{ik}=(q+q^{-1})^{-1}
\sum_{\alpha\neq \beta} \left(
q^{{\rm sign}(\alpha-\beta)} E_{\alpha\alpha}^{(i)} \ot E_{\beta\beta}^{(k)}-
E^{(i)}_{\alpha\beta} \ot E^{(k)}_{\beta\alpha} \right).
\end{equation}
Due to the non-cocommutative coproduct of the quantum group
$U_q(sl(n))$ the q-projectors $P^{\Lambda_2}_{ik}$
are no longer symmetric with respect to $i$ and $k$.
Therefore it is not obvious how $P^{\Lambda_2}_{1N}$ is defined
or how the periodicity for the transfer matrix in eq.~(\ref{1.0})
has to be formulated.
The traditional answer \cite{yang} which means symmetrizing the projectors
as given by eq.~(\ref{1.4}) breaks quantum group invariance,
whereas the projectors given by eq.~(\ref{1.4}) are
quantum group invariant. Therefore one should also have a
quantum group invariant projector $P^{\Lambda_2}_{1N}$.
One possibility to get a quantum group invariant Hamiltonian is to cancel
the projector $P^{\Lambda_2}_{1N}$ in eq.~(\ref{1.2})
which means open boundary conditions, as mentioned above.

For models with periodic boundary conditions
one has to take care of the nontrivial topology of the space, i.e.~of the
graph formed by the square lattice on the cylinder of Fig.~\ref{f1}.
In ref.~\cite{ks} one of the authors of the present paper
and Schrader gave a definition of invariants
of graphs on Riemann surfaces using the language of topological
quantum field theories. The surfaces are considered as boundaries of
3-manifolds.
These invariants are defined for the quantum group case if $q$ is
equal to a root of unity.
We use this definition of invariants of graphs in order to define
the vertex model of eq.~(\ref{1.0}) and Fig.~\ref{f1}.
Using the techniques as formulated in ref.~\cite{ks} we can write
the partition function (\ref{1.0}) in terms of invariants of planar graphs.
As a result we obtain for the transfer matrix of eq.~(\ref{1.0})
as well as for the Hamiltonian eq.~(\ref{1.1}) expressions in
terms of planar graphs which are quantum group invariant and
belong to periodic boundary conditions.
This transition to planar graphs preserves the cyclic invariance
of the models which is obvious from Fig.~\ref{f1}.
It turns out that it is much easier to solve the nested Bethe ansatz
for the periodic case, compared to the open one.
We should add two remarks:
\begin{enumerate}
\item[i.] The invariants of 3-manifolds and therefore also these vertex
models are defined only for quantum groups where $q$ is a root of
unity.\footnote{However, the transfer matrix and the Hamiltonian may
formally be extended to generic values of $q$.}
\item[ii.] The partition function (\ref{1.0}) does not only describe
a two dimensional vertex model on the torus (or cylinder) but in
addition there is a local interaction of the vertices with the interior
of the 3-manifold. However, this interaction is of topological
nature. This model is similar to $\sigma$-models with Chern-Simons
term or the WZNW-models \cite{nov}\cite{wit}.
This will be explained in more details in Appendix A.
\end{enumerate}
In the present paper we solve the eigenvalue equation of the transfer
matrix and the Hamiltonian by means of the algebraic nested Bethe
ansatz and obtain the Bethe ansatz equations. As an application we
calculate the finite size corrections of the ground state energy
using the techniques developed in \cite{karfs} and \cite{karowskieth}.
Therefrom we obtain the central charge of the Virasoro algebra
of the corresponding conformal quantum field theory
%//{1.4a}
\begin{equation}\label{1.4a}
c=(n-1)\left(1-\frac{n(n+1)}{r(r-1)}\right)
\end{equation}
for the $U_q(sl(n))$-model with $q=\exp(i\pi/r),\ (r=n+2,n+3,\dots)$.
This formula coincides with that obtained from the extended coset
construction for $A_{n-1}$ \cite{gko} \cite{bouwknegt}.
It has also been obtained in \cite{japan} for the
$U_q(sl(n))$-RSOS model using Baxter's \cite{baxter}
corner transfer matrix method.
For approaches to quantum group symmetric models with periodic
boundary conditions which use Baxter's SOS-picture of the models
see e.g. refs. \cite{karowskieth}, \cite{japan} and \cite{bazh}.

This paper is organized as follows.
In Section 2 we recall the trigonometric $U_q(sl(n))$-solution of the
Yang-Baxter equation. We use a graphical notation which is useful
in this context to clarify the complicated algebraic structures.
We write some fundamental relations as unitarity, crossing relations,
Markov properties and Cherednik's reflection relation.
In Section 3 we present the transfer matrix of the n-state vertex
model with periodic boundary conditions of
eq.~(\ref{1.0}) and Fig.~\ref{f1}. We define monodromy matrices and
derive commutation rules from the Yang-Baxter relations.
Using these we solve the eigenvalue equation of the transfer matrix
by means of the algebraic nested Bethe ansatz
and obtain the Bethe ansatz equations.
In Section 4 we show how the $U_q(sl(n))$-invariant Hamiltonian (\ref{1.1})
for periodic boundary conditions  is obtained from the transfer matrix.
We perform the finite size analysis of the ground state energy
and obtain the central charge.
In Appendix A we sketch the derivation of the transfer matrix
investigated in this paper. We define the partition function of
the vertex model on a torus using techniques of topological
quantum field theory as developed in ref.~\cite{ks}.
Finally in Appendix B the relation between
Yang-Baxter algebra and quantum groups is used to prove the quantum group
invariance of the transfer matrix and the Hamiltonian.
\section{Yang-Baxter Equation}
Both sides of the Yang-Baxter equation
%//{2.1}
\begin{equation}\label{2.1}
R_{12}(x_{12})R_{13}(x_{13})R_{23}(x_{23})=R_{23}(x_{23})R_{13}(x_{13})
R_{12}(x_{12})
\end{equation}
act on the tensor product space $V_1\otimes V_2\otimes V_3$.
The matrix $R_{ik}(x_{ik})$ depends on the spectral parameter
$x_{ik}=x_i/x_k$ and acts on $V_i\otimes V_k$. The "trigonometric"
$U_q(sl(n))$-solution \cite{babel}, \cite{jimbo}
can be written in terms of the
"constant R-matrix"
%//{2.2}
\begin{equation}\label{2.2}
R(x)=xR-x^{-1}PR^{-1}P
\end{equation}
where $P$ is the permutation operator $P(\alpha\otimes\beta)=
\beta\otimes\alpha$ for $\alpha,\beta\in V$.
The constant R-matrix acting on the tensor product of two fundamental
representation spaces
$V^{\Lambda_1}\otimes V^{\Lambda_1}$ is
%
%//{2.3}
\begin{equation}\label{2.3}
R=R^{\Lambda_1 \Lambda_1}=
\sum_{\alpha\neq \beta}E_{\alpha\alpha}\ot E_{\beta\beta}+q\sum_{\alpha}
E_{\alpha\alpha}\ot E_{\alpha\alpha}+(q-q^{-1})
\sum_{\alpha>\beta}E_{\alpha\beta} \ot E_{\beta\alpha},
\end{equation}
where the $E_{\alpha\beta}$ are the unit matrices in
$M_{n,n}({\bf C})$.
{}From eqs.~(\ref{2.2}) and (\ref{2.3}) one easily derives the relation
%//{2.4}
\begin{equation}\label{2.4}
R(1)=R-PR^{-1}P=(q-q^{-1})P.
\end{equation}

For later convenience we use a graphical notation for matrices
(see e.g. ref.~\cite{resh})
%//{2.5}
\begin{equation}\label{2.5}
A\equiv
\begin{array}{c}
\unitlength=.2mm
\begin{picture}(90.,120.)
\put(10,40){\framebox(80,30){$A$}}
\put(20,25){\line(0,1){15}}
\put(80,25){\line(0,1){15}}
\put(20,70){\line(0,1){15}}
\put(80,70){\line(0,1){15}}
\put(50.,30.){\makebox(0,0)[cc]{$\cdots$}}
\put(50.,80.){\makebox(0,0)[cc]{$\cdots$}}
\put(20.,15.){\makebox(0,0)[cc]{$\scriptstyle\alpha_1$}}
\put(80.,15.){\makebox(0,0)[cc]{$\scriptstyle\alpha_M$}}
\put(20.,100.){\makebox(0,0)[cc]{$\scriptstyle\alpha'_1$}}
\put(80.,100.){\makebox(0,0)[cc]{$\scriptstyle\alpha'_{M'}$}}
\end{picture}
\end{array}
:\left\{
\begin{array}{ccc}
V_1\otimes\cdots\otimes V_M & \to & V'_1\otimes\cdots\otimes V'_{M'}\\
|\alpha_1,\dots,\alpha_M\rangle&\mapsto&|\alpha'_1,\dots,\alpha'_{M'}\rangle
\end{array}
\right.\ .
\end{equation}
For example the matrices $R_{12}$ and $R^{-1}_{12}$:
$V_1\otimes V_2\to V_2\otimes V_1$ defined by eq.~(\ref{2.3})
are depicted by
%//{2.6}
\begin{equation}\label{2.6}
R_{12}\equiv\ \ba{c}
\unitlength=0.3mm
\begin{picture}(30.,30.)
\put(30.,0.){\vector(-1,1){30.}}
\put(18.,18.){\vector(1,1){12.}}
\put(0.,0.){\line(1,1){12.}}
\put(-2.,7.){\makebox(0,0)[cc]{$\scriptstyle 1$}}
\put(32.,7.){\makebox(0,0)[cc]{$\scriptstyle 2$}}
\put(-2.,23.){\makebox(0,0)[cc]{$\scriptstyle 2$}}
\put(32.,23.){\makebox(0,0)[cc]{$\scriptstyle 1$}}
\end{picture}
\ea \qquad \mbox{and} \qquad
R^{-1}_{12}\equiv\ \ba{c}
\unitlength=0.3mm
\begin{picture}(30.,30.)
\put(0.,0.){\vector(1,1){30.}}
\put(12.,18.){\vector(-1,1){12.}}
\put(30.,0.){\line(-1,1){12.}}
\put(-2.,7.){\makebox(0,0)[cc]{$\scriptstyle 1$}}
\put(32.,7.){\makebox(0,0)[cc]{$\scriptstyle 2$}}
\put(-2.,23.){\makebox(0,0)[cc]{$\scriptstyle 2$}}
\put(32.,23.){\makebox(0,0)[cc]{$\scriptstyle 1$}}
\end{picture}
\ea\ .
\end{equation}
The up-arrows denote the representation $\Lambda_1$.
As another example we consider the intertwiners \cite{resh})
%//{2.7}
\begin{equation}\label{2.7}
\ba{c}
V^{\Lambda_1}\ot V^{\Lambda_1^*} \leftrightarrow V^{\Lambda_0}
\qquad\qquad\qquad\qquad
V^{\Lambda_1^*}\ot V^{\Lambda_1} \leftrightarrow V^{\Lambda_0}
\\ \\
\unitlength=0.40mm
\begin{picture}(280.,40.)
\put(75.,25.){\oval(20.,20.)[b]}
\put(85.,35.){\vector(0,-1){5.}}
\put(85.,25.){\line(0,1){5.}}
\put(90.,28.){\makebox(0,0)[lc]{$\equiv q^{(n+1)/2-\alpha}$,}}
\put(15.,17.50){\oval(20.,35.)[t]}
\put(45.,28.){\makebox(0,0)[cc]{or}}
\put(5.,10.){\makebox(0,0)[cc]{$\s \alpha$}}
\put(65.,25.){\vector(0,1){10.}}
\put(230.,25.){\oval(20.,20.)[b]}
\put(175.,17.50){\oval(20.,35.)[t]}
\put(200.,28.){\makebox(0,0)[cc]{or}}
\put(220.,35.){\vector(0,-1){5.}}
\put(220.,25.){\line(0,1){5.}}
\put(240.,25.){\vector(0,1){10.}}
\put(245.,28.){\makebox(0,0)[lc]{$\equiv q^{\alpha-(n+1)/2} $}}
\put(25.,20.){\vector(0,-1){5.}}
\put(5.,15.){\vector(0,1){5.}}
\put(165.,20.){\vector(0,-1){5.}}
\put(185.,15.){\vector(0,1){5.}}
\put(65.,40.){\makebox(0,0)[cc]{$\s \alpha$}}
\put(185.,10.){\makebox(0,0)[cc]{$\s \alpha$}}
\put(240.,40.){\makebox(0,0)[cc]{$\s \alpha$}}
\end{picture}
\ea\qquad .
\end{equation}
The states of $V^{\Lambda_1}$ are labeled by $\alpha\in\{1,2,\dots,n\}$.
The down-arrows denote the conjugate representation $\Lambda_1^*$ and the
trivial representation $\Lambda_0$ is depicted by no line.
{}From the intertwiners (\ref{2.7}) we obtain the "Markov trace"
%
%//{2.8}
\begin{equation}\label{2.8}
\ba{c}
\unitlength=0.50mm
\begin{picture}(20.,35.)
\put(12.50,25.){\oval(15.,20.)[t]}
\put(12.50,10.){\oval(15.,20.)[b]}
\put(20.,25.){\vector(0,-1){15.}}
\put(5.,20.){\makebox(0,0)[cc]{$\s \alpha$}}
\end{picture}
\ea \equiv q^{n+1-2\alpha}
\end{equation}
with the Markov property
%
%//{2.9}
\begin{equation}\label{2.9}
\ba{c}
\unitlength=0.3mm
\begin{picture}(110.,30.)
\put(30.,0.){\vector(-1,1){30.}}
\put(18.,18.){\vector(1,1){12.}}
\put(0.,0.){\line(1,1){12.}}
\put(30.,15.){\oval(30.,30.)[r]}
\put(78.,15.){\makebox(0,0)[cc]{$=q^n$}}
\put(105.,0.){\vector(0,1){30.}}
\end{picture}
\ea \qquad \mbox{and} \qquad
\ba{c}
\unitlength=0.3mm
\begin{picture}(110.,30.)
\put(0.,0.){\vector(1,1){30.}}
\put(12.,18.){\vector(-1,1){12.}}
\put(30.,0.){\line(-1,1){12.}}
\put(30.,15.){\oval(30.,30.)[r]}
\put(80.,15.){\makebox(0,0)[cc]{$=q^{-n}$}}
\put(110.,0.){\vector(0,1){30.}}
\end{picture}
\ea\ ,
\end{equation}
where over the states of the internal lines is summed with the
weights (\ref{2.8}).
In the following summation over internal lines is always assumed.

As another application of the intertwiners (\ref{2.7})
we define R-matrices
acting on other products of $V^{\Lambda_1}$ and $V^{\Lambda_1^*}$
using the R-matrix (\ref{2.6}) and crossing relations, eg.
%
%//{2.10}
\begin{equation}\label{2.10}
R^{\Lambda_1 \Lambda_1^*}=\ba{c}
\unitlength=0.45mm
\begin{picture}(80.,20.)
\put(0.,0.){\line(1,1){8.}}
\put(12.,12.){\vector(1,1){8.}}
\put(0.,20.){\vector(1,-1){20.}}
\put(75.,15.){\oval(10.,10.)[rt]}
\put(80.,15.){\vector(0,-1){15.}}
\put(75.,20.){\line(-1,-1){20.}}
\put(55.,5.){\oval(10.,10.)[lb]}
\put(50.,5.){\line(0,1){15.}}
\put(75.,0.){\line(-1,1){8.}}
\put(63.,12.){\vector(-1,1){8.}}
\put(35.,9.){\makebox(0,0)[cc]{$=$}}
\end{picture}
\ea.
\end{equation}
With these notations the R-matrix inversion and Skein relations
following from eqs.~(\ref{2.3}) and (\ref{2.6}-{\ref{2.10}) read for all
choices of arrows
%
%//{2.11}
\begin{equation}\label{2.11}
\ba{c}
\unitlength=0.45
mm
\begin{picture}(260.,40.)
\put(115.,20.){\line(1,1){8.}}
\put(127.,32.){\line(1,1){8.}}
\put(135.,20.){\line(-1,-1){8.}}
\put(123.,8.){\line(-1,-1){8.}}
\put(145.,20.){\makebox(0,0)[cc]{$+$}}
\put(155.,20.){\line(1,1){20.}}
\put(155.,0.){\line(1,1){20.}}
\put(175.,20.){\line(-1,1){8.}}
\put(155.,20.){\line(1,-1){8.}}
\put(185.,20.){\makebox(0,0)[lc]{$=(q^2+q^{-2})$}}
\put(2.,20.){\line(1,1){8.}}
\put(32.,20.){\makebox(0,0)[rc]{$=$}}
\put(2.,0.){\line(1,1){20.}}
\put(2.,20.){\line(1,-1){8.}}
\put(14.,32.){\line(1,1){8.}}
\put(2.,40.){\line(1,-1){20.}}
\put(22.,0.){\line(-1,1){8.}}
\put(42.,0.){\line(0,1){40.}}
\put(52.,0.){\line(0,1){40.}}
\put(135.,20.){\line(-1,1){20.}}
\put(115.,20.){\line(1,-1){20.}}
\put(175.,0.){\line(-1,1){8.}}
\put(163.,32.){\line(-1,1){8.}}
\put(250.,40.){\line(0,-1){40.}}
\put(260.,0.){\line(0,1){40.}}
\put(260.,40.){\line(0,0){0.}}
\put(85.,20.){\makebox(0,0)[cc]{and}}
\end{picture}
\ea
\quad .
\end{equation}

In addition to this type of graphs considered so far we use graphs
\cite{karthun}
where to each line there belongs not only a representation space
$V$ of $U_q(sl(n))$ but also a "spectral parameter" $x$.
For example the spectral parameter dependent R-matrix given by
eq.~(\ref{2.2}) is denoted by
%
%//{2.12}
\begin{equation}\label{2.12}
R(x/y)=\ba{c}
\unitlength=0.3mm
\begin{picture}(30.,30.)
\put(0.,0.){\vector(1,1){30.}}
\put(30.,0.){\vector(-1,1){30.}}
\put(-5.,0.){\makebox(0,0)[cc]{$\s x$}}
\put(20.,0.){\makebox(0,0)[cc]{$\s y$}}
\end{picture}
\ea=
\frac{x}{y}\ba{c}
\unitlength=0.3mm
\begin{picture}(30.,30.)
\put(30.,0.){\vector(-1,1){30.}}
\put(18.,18){\vector(1,1){12.}}
\put(0.,0.){\line(1,1){12.}}
\end{picture}
\ea -\frac{y}{x} \ba{c}
\unitlength=0.3mm
\begin{picture}(30.,30.)
\put(0.,0.){\vector(1,1){30.}}
\put(12.,18.){\vector(-1,1){12.}}
\put(30.,0.){\line(-1,1){12.}}
\end{picture}
\ea \quad .
\end{equation}
and the graphical notation of the Yang-Baxter equation (\ref{2.1}) is

\bigskip
\parbox{1cm}{}
\hfill
\parbox{10cm}{
$$
\unitlength=.25mm
\begin{picture}(160.,60.)
\put(0,10){\line(1,1){50}}
\put(0,50){\line(1,-1){50}}
\put(38,0){\line(0,1){60}}
\put(110,0){\line(1,1){50}}
\put(110,60){\line(1,-1){50}}
\put(122,0){\line(0,1){60}}
\put(13.,15.){\makebox(0,0)[cc]{$\s 1$}}
\put(31.,6.){\makebox(0,0)[cc]{$\s 2$}}
\put(54.,6.){\makebox(0,0)[cc]{$\s 3$}}
\put(106.,6.){\makebox(0,0)[cc]{$\s 1$}}
\put(129.,6.){\makebox(0,0)[cc]{$\s 2$}}
\put(147.,15.){\makebox(0,0)[cc]{$\s 3$}}
\put(80.,30.){\makebox(0,0)[cc]{=}}
\end{picture}
{}~~.
$$}
\hfill
\parbox{1cm}{
\hfill
(\ref{2.1}')}

\medskip
\noindent
Analogously to eq.~(\ref{2.2}), we have spectral parameter dependent
R-matrices for other representations, e.g.
%//{2.12b}
\beq \label{2.12b}
R^{\Lambda_1 \Lambda_1^*}(x/y)=\frac{x}{y}R^{\Lambda_1 \Lambda_1^*}
-\frac{y}{x}P \left( R^{\Lambda_1^* \Lambda_1} \right)^{-1} P.
\eeq
As a spectral parameter dependent version of eq.~(\ref{2.10})
we have crossing relations like
%
%//{2.12c}
\begin{equation}\label{2.12c}
\ba{c}
\unitlength=0.50mm
\begin{picture}(84.,20.)
\put(-3.,2.){\makebox(0.,0.){$\s x$}}
\put(23.,2.){\makebox(0.,0.){$\s y$}}
\put(61.,0.){\makebox(0.,0.){$\s -y$}}
\put(70.,0.){\makebox(0.,0.){$\s x$}}
\put(84.,2.){\makebox(0.,0.){$\s y$}}
\put(0.,0.){\line(1,1){20.}}
\put(0.,20.){\line(1,-1){20.}}
\put(75.,15.){\oval(10.,10.)[rt]}
\put(80.,15.){\line(0,-1){15.}}
\put(75.,20.){\line(-1,-1){20.}}
\put(55.,5.){\oval(10.,10.)[lb]}
\put(50.,5.){\line(0,1){15.}}
\put(75.,0.){\line(-1,1){20.}}
\put(35.,9.){\makebox(0,0)[cc]{$=$}}
\end{picture}
\ea
\end{equation}
for all choices of arrows, if we introduce, as an extension of
rel. (\ref{2.7}), spectral parameter dependent intertwiners by
%//{2.12d}
\beq \label{2.12d}
\ba{c}
\unitlength=.6mm
\begin{picture}(115.,15.)
\put(27.,0.){\oval(15.,20.)[t]}
\put(97.,10.){\oval(15.,20.)[b]}
\put(63.,5.){\makebox(0,0){and}}
\put(15.,2.){\makebox(0,0){$\s x$}}
\put(41.,2.){\makebox(0,0){$\s -x$}}
\put(84.,8.){\makebox(0,0){$\s x$}}
\put(111.,8.){\makebox(0,0){$\s -x$}}
\end{picture}
\ea\ ,\eeq
where the spectral parameter $x$ changes sign.
Using again intertwiners and crossing relations one derives
from eq.~(\ref{2.12}) for all choices of arrows the inversion relation for
the spectral parameter dependent R-matrix
%//{2.13}
\begin{equation}\label{2.13}
\ba{c}
\unitlength=0.4mm
\begin{picture}(25.,40.)
\put(5.,0.){\line(1,1){20.}}
\put(25.,20.){\line(-1,1){20.}}
\put(25.,40.){\line(-1,-1){20.}}
\put(5.,20.){\line(1,-1){20.}}
\put(25.,5.){\makebox(0,0)[cc]{$\s y$}}
\put(5.,5.){\makebox(0,0)[cc]{$\s x$}}
\end{picture}
\ea =(q^2+q^{-2}-x^2/y^2-y^2/x^2)
\ba{c}
\unitlength=0.5mm
\begin{picture}(25.,40.)
\put(5.,0.){\line(0,1){40.}}
\put(20.,0.){\line(0,1){40.}}
\put(0.,5.){\makebox(0,0)[cc]{$\s x$}}
\put(25.,5.){\makebox(0,0)[cc]{$\s y$}}
\end{picture}
\ea\ .
\end{equation}

In addition we will make use of Cherednik's \cite{cher} reflection
property. We only need the case of the reflection matrix $K={\bf 1}$.
For all choices of arrows one derives from eqs.~(\ref{2.11}) and (\ref{2.12})
%//{2.15}
\begin{equation}\label{2.15}
\ba{c}
\unitlength=0.50mm
\begin{picture}(100.,70.)
\put(20.,10.){\line(-1,2){10.}}
\put(10.,45.){\line(3,-1){30.}}
\put(94.,12.){\makebox(0,0)[cc]{$\s x$}}
\put(85.,65.){\makebox(0,0)[cc]{$\s \mu /x$}}
\put(35.,30.){\makebox(0,0)[cc]{$\s y$}}
\put(100.,40.){\makebox(0,0)[cc]{$\s \mu /y$}}
\put(55.,40.){\makebox(0,0)[cc]{$=$}}
\put(70.,50.){\line(1,-2){20.}}
\put(70.,35.){\line(3,-1){30.}}
\put(95.,20.){\makebox(0,0)[cc]{$\s y$}}
\put(24.,12.){\makebox(0,0)[cc]{$\s x$}}
\put(38.,54.){\makebox(0,0)[lc]{$\s \mu /y$}}
\put(40.,67.){\makebox(0,0)[lc]{$\s \mu /x$}}
\put(10.,10.){\dashbox{1.}(0.,60.)[cc]{}}
\put(70.,10.){\dashbox{1.}(0.,60.)[cc]{}}
\put(70.,50.){\line(1,2){10.}}
\put(10.,30.){\line(3,4){30.}}
\put(10.,45.){\line(2,1){30.}}
\put(70.,35.){\line(3,1){30.}}
\end{picture}
\ea
\end{equation}
where the spectral parameter at the reflection point (at the dotted line,
later denoted by a bar) changes from $x$ to $\mu /x$ and $y$ to $\mu/y$
for an arbitrary constant $\mu$. In Sect. 3 we will make use of this
arbitrariness and take the limit $\mu\to\infty$ which simplifies
the nested Bethe ansatz.
\section{The algebraic nested Bethe ansatz}
In order to solve the eigenvalue equation for the Hamiltonian (\ref{1.1})
we introduce the n-state vertex model of eq.~(\ref{1.0}).
As usual \cite{fad} we introduce a monodromy matrix as a product
of spectral parameter dependent R-matrices as follows
%
%//{3.1}
\begin{equation}\label{3.1}
T_\alpha^\beta(x,\{x_i\}):=\left[
R(x_N/x)\ldots R(x_2/x)R(x_1/x)\right]^\beta_\alpha=
\ba{c}
\unitlength=0.50mm
\begin{picture}(99.,29.)
\put(85.,15.){\line(1,-1){10.}}
\put(20.,15.){\vector(-1,1){10.}}
\put(60.,15.){\line(1,0){25.}}
\put(50.,15.){\makebox(0,0)[cc]{$\cdots$}}
\put(35.,5.){\makebox(0,0)[cc]{$\s x_N$}}
\put(70.,5.){\makebox(0,0)[cc]{$\s x_2$}}
\put(80.,5.){\makebox(0,0)[cc]{$\s x_1$}}
\put(95.,10.){\makebox(0,0)[cc]{$\s x$}}
\put(20.,15.){\line(1,0){20.}}
\put(99.,2.){\makebox(0,0)[cc]{$\s \alpha$}}
\put(10.,29.){\makebox(0,0)[rc]{$\s \beta$}}
\put(30.,0.){\vector(0,1){25.}}
\put(65.,0.){\vector(0,1){25.}}
\put(75.,0.){\vector(0,1){25.}}
\end{picture}
\ea
\eeq
where to all lines (the horizontal one and the vertical ones)
the fundamental representation $\Lambda_1$ is associated.
We have omitted the indices which belong to the vertical space
%
%//{3.2}
\begin{equation}\label{3.2}
\Omega=V_N^{\Lambda_1} \ot \ldots \ot V_1^{\Lambda_1} .
\end{equation}
For the rational case $q=1$, one obtains an $SU(n)$-invariant
transfer matrix for periodic boundary conditions
as the trace over the horizontal space of the monodromy
matrix (\ref{3.1}): $\tau|_{q=1}=\sum_{\alpha=1}^n T_\alpha^\alpha$.
For the quantum group case this trace does not yield an
$U_q(sl(n))$-invariant transfer matrix. However, the transfer matrix of
eq.~(\ref{1.0}) associated to the vertex model depicted in Fig.~\ref{f1}
corresponds to periodic boundary conditions and should be an $U_q(sl(n))$
invariant. We show in Appendix A
using the techniques of topological quantum field theory
developed in \cite{ks} that this transfer matrix
is given by the graph
%
%//{3.3}
\vskip2mm
\begin{equation}\label{3.3}
\tau^{\underline\alpha'}_{\underline\alpha}(x,\{x_i\})=\ba{c}
\unitlength=0.50mm
\begin{picture}(100.,50.)
\put(55.,15.){\line(1,0){25.}}
\put(40.,15.){\makebox(0,0)[cc]{$\cdots$}}
\put(25.,5.){\makebox(0,0)[cc]{$\s x_N$}}
\put(65.,5.){\makebox(0,0)[cc]{$\s x_2$}}
\put(75.,5.){\makebox(0,0)[cc]{$\s x_1$}}
\put(20.,-4.){\makebox(0,0)[cc]{$\s \alpha_N$}}
\put(60.,-4.){\makebox(0,0)[cc]{$\s \alpha_2$}}
\put(70.,-4.){\makebox(0,0)[cc]{$\s \alpha_1$}}
\put(20.,50.){\makebox(0,0)[cc]{$\s \alpha'_N$}}
\put(60.,50.){\makebox(0,0)[cc]{$\s \alpha'_2$}}
\put(70.,50.){\makebox(0,0)[cc]{$\s \alpha'_1$}}
\put(85.,15.){\makebox(0,0)[lb]{$\s x$}}
%\put(20.,15.){\oval(10.,10.)[rb]}
%\put(60.,15.){\oval(10.,10.)[rb]}
%\put(70.,15.){\oval(10.,10.)[rb]}
\put(15.,25.){\oval(20.,20.)[l]}
\put(70.,0.){\vector(0,1){45.}}
\put(60.,0.){\vector(0,1){45.}}
\put(20.,0.){\vector(0,1){45.}}
\put(50.,35.){\line(1,0){5.}}
\put(50.,15.){\line(1,0){5.}}
\put(40.,35.){\makebox(0,0)[cc]{$\cdots$}}
\put(5.,25.){\makebox(0,0)[cc]{-}}
\put(22.,35.){\line(1,0){8.}}
\put(50.,35.){\line(1,0){8.}}
\put(62.,35.){\line(1,0){6.}}
\put(72.,35.){\line(1,0){8.}}
\put(80.,35.){\line(0,0){0.}}
\put(80.,40.){\oval(10.,10.)[rb]}
\put(80.,10.){\oval(10.,10.)[rt]}
\put(92.50,40.){\oval(15.,20.)[t]}
\put(92.50,10.){\oval(15.,20.)[b]}
\put(100.,25.){\makebox(0,0)[cc]{-}}
\put(30.,15.){\vector(-1,0){16.}}
\put(14.,35.){\line(1,0){4.}}
\put(74.,35.){\vector(1,0){4.}}
\put(100.,41.){\line(0,-1){12.}}
\put(100.,9.){\line(0,1){22.}}
\end{picture}
\ea,
\end{equation}
\vskip2mm\noindent
where in the lower row the monodromy matrix (\ref{3.1}), in the upper one a
product of constant R-matrices and on the right the Markov trace
(\ref{2.8}) appear.
As in eq.~(\ref{2.15})  the two bars denote the transition of the spectral
parameter from $x$ to $x'=\mu/x=\infty$ (note that
$R(x'\to\infty)\approx x'R$). Obviously, for the rational case
$q=1$ the transfer matrix (\ref{3.3}) agrees with the conventional one
mentioned
above, since for $q=1$ the Markov trace as well as the R-matrix
($R|_{q=1}={\bf 1}$) are trivial.

In order to perform the algebraic (nested) Bethe ansatz for the
transfer matrix (\ref{3.3})
we introduce two more monodromy matrices
%
%//{3.4}
\begin{equation}\label{3.4}
\tilde{T}^\beta_\alpha(x,\{x_i\})=
\ba{c}
\unitlength=0.50mm
\begin{picture}(89.,34.)
\put(25.,10.){\vector(0,1){20.}}
\put(55.,10.){\vector(0,1){20.}}
\put(65.,10.){\vector(0,1){20.}}
\put(50.,20.){\line(1,0){25.}}
\put(40.,20.){\makebox(0,0)[cc]{$\cdots$}}
\put(30.,10.){\makebox(0,0)[cc]{$\s x_N$}}
\put(60.,10.){\makebox(0,0)[cc]{$\s x_2$}}
\put(70.,10.){\makebox(0,0)[cc]{$\s x_1$}}
\put(30.,20.){\line(-1,0){15.}}
\put(5.,10.){\line(1,1){10.}}
\put(75.,20.){\vector(1,1){10.}}
\put(12.,12.){\makebox(0,0)[cc]{$\s x$}}
\put(89.,34.){\makebox(0,0)[cc]{$\s \beta$}}
\put(1.,6.){\makebox(0,0)[cc]{$\s \alpha$}}
\put(25.,5.){\line(0,1){7.}}
\put(55.,5.){\line(0,1){7.}}
\put(65.,5.){\line(0,1){6.}}
\end{picture}
\ea,
\end{equation}
and
%//{3.5}
\begin{equation}\label{3.5}
\tmod^\beta_\alpha(x,\{x_i\};\mu):=\sum_{\ga=1}^n
\tilde{T}^\beta_\ga(\mu/x,\{x_i\})
T^\ga_\alpha(x,\{x_i\})
=\ba{c}
\unitlength=0.50mm
\begin{picture}(95.,49.)
\put(80.,15.){\line(1,-1){10.}}
\put(55.,15.){\line(1,0){25.}}
\put(40.,15.){\makebox(0,0)[cc]{$\cdots$}}
\put(25.,5.){\makebox(0,0)[cc]{$\s x_N$}}
\put(65.,5.){\makebox(0,0)[cc]{$\s x_2$}}
\put(75.,5.){\makebox(0,0)[cc]{$\s x_1$}}
\put(90.,10.){\makebox(0,0)[cc]{$\s x$}}
\put(94.,2.){\makebox(0,0)[cc]{$\s \alpha$}}
\put(15.,25.){\oval(20.,20.)[l]}
\put(80.,35.){\vector(1,1){10.}}
\put(70.,0.){\vector(0,1){45.}}
\put(60.,0.){\vector(0,1){45.}}
\put(20.,0.){\vector(0,1){45.}}
\put(55.,35.){\line(1,0){25.}}
\put(50.,35.){\line(1,0){5.}}
\put(50.,15.){\line(1,0){5.}}
\put(40.,35.){\makebox(0,0)[cc]{$\cdots$}}
\put(5.,25.){\makebox(0,0)[cc]{-}}
\put(95.,49.){\makebox(0,0)[cc]{$\s \beta$}}
\put(4.,34.){\makebox(0,0)[rc]{$\s \mu/x$}}
\put(15.,15.){\line(1,0){15.}}
\put(15.,35.){\line(1,0){15.}}
\end{picture}
\ea\ ,
\end{equation}
where the bar again denotes the transition of the spectral parameter
$x$ to $\mu/x$ as in eq.~(\ref{2.15}).
(Note that because of eq.~(\ref{2.13}) $\tilde T\propto T^{-1}$.)
The monodromy matrix (\ref{3.5}) has been used in refs.~\cite{skly}
\cite{devega2q}\cite{deveganq}\cite{forster}
for the case $\mu=1$ which corresponds to open boundary conditions.
We shall see below that the nested Bethe ansatz simplifies drastically
for $\mu\to\infty$. This corresponds to periodic boundary conditions.

The Yang-Baxter equation (\ref{1.1}) and the reflection property
(\ref{2.15}) imply the following Yang-Baxter relation for the monodromy matrix
$\cal T$ given by eq.~(\ref{3.5}):
%
%//{3.7}
\beq \label{3.7}
R^{\alpha\beta}_{\alpha'\beta'}(y/x)\tmod^{\beta'}_{\ga'}(x;\mu)R^{\ga' \de}_
{\ga \de'}(\mu/xy)\tmod^{\de'}_\de(y;\mu)=
\tmod^\alpha_{\alpha''}(y;\mu)R^{\alpha'' \beta}_{\de''\beta''}(\mu/xy)
\tmod^{\beta''}_{\ga''}(x;\mu)
R^{\ga'' \de''}_{\ga\de}(y/x) ,
\eeq
graphically expressed by
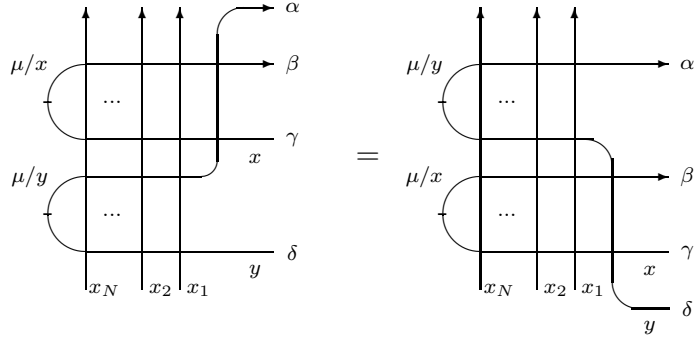
\[
\unitlength=0.50mm
\begin{picture}(170.,85.)
\put(10.,30.){\oval(20.,20.)[l]}
\put(10.,60.){\oval(20.,20.)[l]}
\put(55.,45.){\makebox(0,0)[cc]{$\s x$}}
\put(55.,15.){\makebox(0,0)[cc]{$\s y$}}
\put(115.,30.){\oval(20.,20.)[l]}
\put(115.,60.){\oval(20.,20.)[l]}
\put(115.,40.){\line(1,0){25.}}
\put(115.,70.){\line(1,0){25.}}
\put(115.,50.){\line(1,0){25.}}
\put(115.,20.){\line(1,0){25.}}
\put(85.,45.){\makebox(0,0)[cc]{$=$}}
\put(0.,60.){\makebox(0,0)[cc]{-}}
\put(0.,30.){\makebox(0,0)[cc]{-}}
\put(105.,60.){\makebox(0,0)[cc]{-}}
\put(105.,30.){\makebox(0,0)[cc]{-}}
\put(10.,10.){\vector(0,1){75.}}
\put(35.,10.){\vector(0,1){75.}}
\put(25.,10.){\vector(0,1){75.}}
\put(115.,10.){\vector(0,1){75.}}
\put(140.,10.){\vector(0,1){75.}}
\put(130.,10.){\vector(0,1){75.}}
\put(40.,10.){\makebox(0,0)[cc]{$\s x_1$}}
\put(15.,10.){\makebox(0,0)[cc]{$\s x_N$}}
\put(30.,10.){\makebox(0,0)[cc]{$\s x_2$}}
\put(145.,10.){\makebox(0,0)[cc]{$\s x_1$}}
\put(120.,10.){\makebox(0,0)[cc]{$\s x_N$}}
\put(135.,10.){\makebox(0,0)[cc]{$\s x_2$}}
\put(17.,30.){\makebox(0,0)[cc]{$\s \ldots$}}
\put(15.,70.){\vector(1,0){45.}}
\put(60.,50.){\line(-1,0){45.}}
\put(55.,20.){\line(-1,0){45.}}
\put(10.,40.){\line(1,0){25.}}
\put(40.,50.){\oval(10.,20.)[rb]}
\put(52.50,67.50){\oval(15.,35.)[lt]}
\put(45.,70.){\line(0,-1){20.}}
\put(50.,85.){\vector(1,0){10.}}
\put(142.50,42.50){\oval(15.,15.)[rt]}
\put(157.50,20.){\oval(15.,30.)[lb]}
\put(145.,70.){\vector(1,0){20.}}
\put(150.,45.){\line(0,-1){30.}}
\put(145.,40.){\vector(1,0){20.}}
\put(165.,20.){\line(-1,0){20.}}
\put(155.,5.){\line(1,0){10.}}
\put(160.,0.){\makebox(0,0)[cc]{$\s y$}}
\put(160.,15.){\makebox(0,0)[cc]{$\s x$}}
\put(65.,85.){\makebox(0,0)[cc]{$\s \alpha$}}
\put(65.,70.){\makebox(0,0)[cc]{$\s \beta$}}
\put(65.,50.){\makebox(0,0)[cc]{$\s \ga$}}
\put(65.,20.){\makebox(0,0)[cc]{$\s \de$}}
\put(170.,70.){\makebox(0,0)[cc]{$\s \alpha$}}
\put(170.,40.){\makebox(0,0)[cc]{$\s \beta$}}
\put(170.,20.){\makebox(0,0)[cc]{$\s \ga$}}
\put(170.,5.){\makebox(0,0)[cc]{$\s \de$}}
\put(105.,70.){\makebox(0,0)[rc]{$\s \mu/y$}}
\put(105.,40.){\makebox(0,0)[rc]{$\s \mu/x$}}
\put(0.,70.){\makebox(0,0)[rc]{$\s \mu/x$}}
\put(0.,40.){\makebox(0,0)[rc]{$\s \mu/y$}}
\put(17.,60.){\makebox(0,0)[cc]{$\s \ldots$}}
\put(122.,30.){\makebox(0,0)[cc]{$\s \ldots$}}
\put(122.,60.){\makebox(0,0)[cc]{$\s \ldots$}}
\put(140.,70.){\line(1,0){5.33}}
\put(145.,50.){\line(-1,0){5.}}
\put(145.,40.){\line(-1,0){5.}}
\put(140.,20.){\line(1,0){5.33}}
\put(40.,70.){\line(1,0){5.33}}
\put(40.,40.){\line(-1,0){4.67}}
\put(10.,50.){\line(1,0){5.}}
\put(10.,70.){\line(1,0){5.}}
\put(55.,20.){\line(1,0){5.}}
\end{picture}
\]
Compared to the case of the monodromy matrix $T$
these commutation relations are more complicated.
The usual decomposition into "wanted" and "unwanted"
terms does not appear (see refs.~\cite{deveganq}\cite{forster}).
We obtain much simpler commutation rules in the limit $\mu\to\infty$.
The contributions from $\tilde T$ reduce to
constant R-matrices. Also two R-matrices in eq.~(\ref{3.7}) become constant.
So the commutation relations simplify in an essential way.

We write the monodromy matrix $\cal T$ (up to a factor) in the limit
$\mu\to\infty$ in block form as a matrix in the horizontal space
%
%//{3.8}
\begin{equation}\label{3.8}
\tmod(x)=\left( \ba{ll} \amod(x) & \bemod(x) \\\cmod(x) & \dmod(x)
\ea \right)
=\lim_{\mu\to \infty} \Big(\prod_{i=1}^{N}\frac{xx_i}{\mu}\Big)\tmod(x;\mu)
\end{equation}
where
$\amod \in  M_{1,1}( End(\Omega))$, $\bemod \in M_{1,n-1}( End(\Omega))$,
etc.. From eq.~(\ref{3.7}) we get the commutation relations of the
operator valued entries of $\cal T$
%//{3.9a}
\begin{equation}\label{3.9a}
\amod(x)\bemod_\gamma(y)=\underbrace{q^{-1}a(x/y)\bemod_\ga(y)\amod(x)}
_{\mbox{wanted term}}-\underbrace{
q^{-1}\{c_-(x/y)\bemod_\ga(x)\amod(y)+(q-q^{-1})\bemod_\alpha
(x)\dmod^\alpha_\ga(y)\} }_{\mbox{unwanted term}}\quad,
\eeq
%//{3.9b}
\beq \label{3.9b}
\dmod^\beta_\ga(x)\bemod_\de(y)=
\underbrace{\bemod_{\beta'}(y)\dmod^\alpha_{\ga'}(x)
{R}^{\beta' \beta}_{\de'\alpha} R^{\ga'\de'}_{\ga\de}(y/x)}
_{\mbox{wanted term}}-\underbrace{
c_-(y/x)\bemod_{\beta''}(x)\dmod^{\alpha'}_\de(y)
{R}^{\beta''\beta}_{\ga\alpha'}}_{\mbox{unwanted term}}\quad,
\end{equation}
where the entries $a(x)$ and $c_-(x)$ of $R(x)$ are obtained from
eqs.~(\ref{2.3}) and (\ref{2.12})
%//{3.10}
\begin{equation}\label{3.10}
a(x)=\frac{x^2q-q^{-1}}{x^2-1}\quad \mbox{and} \quad
c_-(x)=-\frac{q-q^{-1}}{x^2-1}\ .
\end{equation}

The transfer matrix (\ref{3.3}) is the Markov trace (\ref{2.8}) of the
monodromy matrix $\cal T$ defined by eq.~(\ref{3.8})
%
%//{3.6}
\begin{equation}\label{3.6}
\tau=\tr_q \tmod=\sum_{\alpha=1}^n \tmod^\alpha_\alpha
q^{n+1-2\alpha}
=q^{n-1}\amod+\sum_{\alpha=2}^n q^{n+1-2\alpha} \dmod.
\end{equation}
Using crossing, inversion and reflection relations (\ref{2.10},
\ref{2.13} and \ref{2.15})
one proves that eq.~(\ref{3.7}) implies that transfer matrices
(\ref{3.6}) commute for different spectral parameters. In Appendix B we
show in addition that the transfer matrix commutes with the generators of the
quantum group $U_q(sl(n))$. These generators are derived from the monodromy
matrices $T$ or $\tmod$ in the limits $x\to 0$ or $\infty$.

We now apply the algebraic nested Bethe ansatz method to the eigenvalue
problem of the transfer matrix
%//{3.11}
\begin{equation}\label{3.11}
\tau(x)\Psi=\Lambda(x)\Psi
\end{equation}
for $\Psi \in \Omega$.
The action of $\cal T$ on the reference state
$\Phi:=\bigotimes_{i=1}^N |1\rangle_i \in \Omega$
which is one of the ferromagnetic ground states is given by
%//{3.12}
\begin{eqnarray}\label{3.12}
\amod(x,\{x_i\})\Phi&=&
q^N \prod_{i=1}^N a(x_i/x) \Phi \nonumber \\
\dmod^\alpha_\beta(x,\{x_i\})\Phi&=&\de^\alpha_\beta \prod_{i=1}^N \Phi\\
\bemod_\beta(x,\{x_i\})\Phi&\neq& 0,\ \not\propto \Phi \nonumber \\
\cmod^\alpha(x,\{x_i\})\Phi&=&0.\nonumber
\end{eqnarray}
We construct a Bethe ansatz vector
by repeated action ($r$ times) of creation operators $\bemod$ on $\Phi$
\beq
\Psi(\{\hat{x_i}\}):=\left\{ \sum_{\e_1\ldots \e_r=2}^n
\bemod_{\e_1}(\hat{x}_1)
\ldots \bemod_{\e_r}(\hat{x}_r)\hat{\Psi}^{\e_1\ldots \e_r}\right\}\Phi\quad,
\eeq
where $\hat{\Psi}$ is an element of the reduced quantum space
$\hat{\Omega}$ representing a chain of length $r$ and
admitting only states $|2\rangle,\ldots ,|n\rangle$.
To compute the action of $\amod$ and $\dmod$ in the eigenvalue eq.
(\ref{3.11}) we commute them through all $\bemod$'s to the right
and apply them to $\Phi$.
The wanted terms in eqs.~(\ref{3.9a}) and (\ref{3.9b})
together with (\ref{3.12}) yield
%//{3.14}
\begin{eqnarray} \label{3.14}
q^{n-1}\amod(x)\Psi
&=&q^{n-1}\amod(x)\left\{ \sum_{\e_1\ldots \e_r=2}^n
\bemod_{\e_1}(\hat{x}_1)\ldots \bemod_{\e_r}(\hat{x}_r)\hat{\Psi}^{\e_1\ldots
\e_r}\right\}\Phi\nonumber\\
&=&q^{N-r+n-1}\prod_{i=1}^r a\big(x/\hat{x}_i\big)\prod_{j=1}^N
a\big(x_j/x\big)\Psi+\mbox{uwt}
\end{eqnarray}
and
%//{3.15}
\begin{eqnarray}\label{3.15}
\lefteqn{\sum_{\alpha=2}^nq^{n+1-2\alpha}\dmod^\alpha_\alpha(x)\Psi
=\sum_{\alpha=2}^nq^{n+1-2\alpha}\dmod^\alpha_\alpha(x)\left\{
\sum_{\e_1\ldots \e_r=2}^n
\bemod_{\e_1}(\hat{x}_1)\ldots \bemod_{\e_r}(\hat{x}_r)\hat{\Psi}^{\e_1\ldots
\e_r}\right\}\Phi}\nonumber\\
&=&\sum_{\e_1\ldots \e_r=2}^n \bemod_{\beta_1}(\hat{x}_1)\ldots
\bemod_{\beta_r}(\hat{x}_r)\nonumber\\
&&\times\left\{ \sum_{\alpha=2}^nq^{n+1-2\alpha}
{R}^{\beta_r\alpha_{r-1}}_{\de_r\alpha_r}R^{\ga_r\de_r}_{\ga_{r-1}\e_r}
(\hat{x_r}/x)\ldots {R}^{\beta_1\alpha}_{\de_1\alpha_1}R^{\ga_1\de_1}
_{\alpha \e_1}(\hat{x_r}/x)
\hat{\Psi}^{\e_1\ldots \e_r}\right\}\dmod^{\alpha_r}_{\ga_r}\Phi+\mbox{uwt}
\nonumber\\
&=&\prod_{\e_1\ldots \e_r=2}^n \bemod_{\beta_1}(\hat{x}_1)\ldots
\bemod_{\beta_r}(\hat{x}_r)\left\{q^{-1} \hat{\tau} \hat{\Psi}
\right\}\Phi+\mbox{uwt},
\end{eqnarray}
where we have introduced an $U_q(sl(n-1))$ monodromy matrix $\hat{\cal T}$ and
a transfer matrix
%//{3.16}
\begin{equation}\label{3.16}
\hat\tau=\sum_{\alpha=1}^{n-1}\hat{\tmod}^\alpha_\alpha q^{n+1-2\alpha}.
\end{equation}
The first term (wanted term) on the right hand side of eq.~(\ref{3.15})
is proportional to $\Psi$ if the eigenvalue equation
%//{3.17}
\begin{equation}\label{3.17}
\hat\tau \hat{\Psi}=\hat{\Lambda}\hat{\Psi}
\end{equation}
is fulfilled. This problem agrees with the original one (\ref{3.11})
where $n$ is replaced by $n-1$. Thus iterating the procedure
described above from level $n$ to $1$ we solve the eigenvalue
equation (\ref{3.11}). In the following,
all operators, Bethe ansatz parameters etc. will be labeled
by the number of the corresponding Bethe ansatz level, in particular
$r_n=N,\ r_{n-1}=r,\ r_0=0,\ x_i^{(n)}=x_i,\ x_i^{(n-1)}=\hat x_i$.

In case all unwanted terms cancel, the eigenvalue equation of the transfer
matrix eq.~(\ref{3.11}) is solved and the eigenvalue $\Lambda(x)$
consists of the wanted coefficients $\lambda_k$
%
%//{3.18}
\begin{equation}\label{3.18}
\Lambda(x) =\sum_{k=1}^n \lambda_k(x),
\end{equation}
where
%//{3.19}
\beq\label{3.19}
\lambda_k(x)=q^{2k-n-1+r_{k}-r_{k-1}}\prod_{i=1}^{r_{k-1}}
a\big({x/x_i^{(k-1)}}\big)\prod_{j=1}^{r_k}
a\big({x_j^{(k)}/x}\big),\qquad (k=1,\ldots,n).
\eeq
The Bethe ansatz equations are equivalent to the vanishing of all
unwanted terms. They
can be obtained from the property that $\Lambda(x)$
must have finite values when the spectral parameter $x$ approaches one of the
Bethe ansatz parameters $x_i^{(k)}$, because $\tau(x)$ is an analytical
function in $x$. Writing $x_m^{(k)}=\exp \th_m^{(k)}$ and $q=\exp i\ga$
we obtain the coupled system of Bethe ansatz equations:
%//{3.20}
\begin{eqnarray}  \label{3.20}
\lefteqn{ q^{2+\eta_{n-k}}
\prod_{i_1=1}^{r_k}\frac{\sinh\left(\th_m^{(k)}-\th_{i_1}^{(k)}+i\ga\right)}
{\sinh\left(\th_m^{(k)}-\th_{i_1}^{(k)}-i\ga\right)}=
-\prod_{i_2=1}^{r_{k+1}}\frac{\sinh\left(\th_m^{(k)}-
\th_{i_2}^{(k+1)}\right)}
{\sinh\left(\th_m^{(k)}-\th_{i_2}^{(k+1)}-i\ga\right)}} \nonumber \\
&&\qquad \qquad \qquad \qquad \times
\prod_{i_3=1}^{r_{k-1}}\frac{\sinh\left(\th_m^{(k)}-\th_{i_3}^{(k-1)}+
i\ga\right)}
{\sinh\left(\th_m^{(k)}-\th_{i_3}^{(k-1)}\right)}\ ,\quad (k=1,\ldots,n-1),
\end{eqnarray}
where $\eta_{n-k}=r_{k+1}-2r_k+r_{k-1}$ are the eigenvalues of the $U_q(sl(n))$
Cartan elements $H_{n-k},\ (k=1,\ldots,n-1)$ of the Bethe ansatz
vector $\Psi$ (see Appendix B).
\section{The Hamiltonian and finite size analysis}
The Hamiltonian (\ref{1.1}) may be obtained from the transfer matrix
(\ref{3.3}) as follows
%//{4.1}
\beq \label{4.1}
H={\rm const.}
\left. \frac{d}{dx}\ln \tau(x,\{x_i\}\right|_{x=x_i=1}={\rm const.}
\left(
\sum_{i=1}^{N-1}
\ba{c}
\unitlength=0.50mm
\begin{picture}(100.,45.)
\put(5.,10.){\vector(0,1){30.}}
\put(10.,25.){\makebox(0,0)[cc]{$\s \ldots$}}
\put(15.,15.){\line(1,2){10.}}
\put(25.,15.){\line(-1,2){10.}}
\put(15.,35.){\vector(0,1){5.}}
\put(25.,35.){\vector(0,1){5.}}
\put(15.,10.){\line(0,1){5.}}
\put(25.,10.){\line(0,1){5.}}
\put(30.,25.){\makebox(0,0)[cc]{$\s \ldots$}}
\put(35.,10.){\vector(0,1){30.}}
\put(20.,25.){\makebox(0,0)[cc]{$\bullet$}}
\put(54.,25.){\makebox(0,0)[cc]{$+$}}
\put(80.,35.){\vector(0,1){5.}}
\put(80.,10.){\line(0,1){5.}}
\put(90.,25.){\makebox(0,0)[cc]{$\s \ldots$}}
\put(95.,10.){\vector(0,1){30.}}
\put(75.,25.){\makebox(0,0)[cc]{$\bullet$}}
\put(85.,10.){\vector(0,1){30.}}
\put(95.,5.){\makebox(0,0)[cc]{$\s 2$}}
\put(80.,5.){\makebox(0,0)[cc]{$\s N$}}
\put(35.,5.){\makebox(0,0)[cc]{$\s 1$}}
\put(5.,5.){\makebox(0,0)[cc]{$\s N$}}
\put(15.,5.){\makebox(0,0)[cc]{$\s i+1$}}
\put(25.,5.){\makebox(0,0)[cc]{$\s i$}}
\put(100.,5.){\makebox(0,0)[cc]{$\s 1$}}
\put(70.,20.){\line(1,1){10.}}
\put(80.,20.){\line(-1,1){10.}}
\put(75.,30.){\oval(10.,10.)[lt]}
\put(75.,20.){\oval(10.,10.)[lb]}
\put(80.,30.){\line(0,1){5.}}
\put(80.,15.){\line(0,1){5.}}
\put(97.50,40.){\oval(5.,10.)[rb]}
\put(97.50,10.){\oval(5.,10.)[rt]}
\put(100.,39.){\vector(0,1){1.}}
\put(93.,35.){\line(-1,0){2.}}
\put(87.,35.){\line(1,0){5.}}
\put(87.,15.){\line(1,0){6.}}
\put(78.,35.){\line(-1,0){3.}}
\put(78.,15.){\line(-1,0){3.}}
\put(82.,35.){\line(1,0){1.}}
\put(83.,15.){\line(-1,0){1.}}
\end{picture}
\ea
\right)\ .
\eeq
The dotted crossings mean the matrix (see eq.~(\ref{2.2}))
%//{4.3}
\beq \label{4.3}
\bigg(R^{-1}(x)\frac{d}{dx}R(x)\bigg)_{x=1}=\frac{1}{q-q^{-1}}(PR+R^{-1}P)
=\frac{q+q^{-1}}{q-q^{-1}}\big({\bf 1} -2P^{\Lambda_2}\big),
\eeq
where the following decomposition (\ref{1.3}) of the R-matrix \cite{jimbo3}
and the completeness of the projectors has been used
%
%//{4.2}
\beq \label{4.2}
PR=qP^{2\Lambda_1}-q^{-1}P^{\Lambda_2}.
\eeq
Therefore we get (with ${\rm const.}=-1/2(q-q^{-1})/(q+q^{-1})$)
up to terms proportional to the unity operator the Hamiltonian of
eq.~(\ref{1.1}) as a sum of projectors $P^{\Lambda_2}$.

In Appendix A we derive the transfer matrix $\tau$ from a cyclic invariant
vertex model on a torus. Therefrom it it obvious that the transfer matrix
as well as the Hamiltonian (\ref{4.1}) describe models with periodic
boundary conditions. However, the derivation in Appendix~A
relies on methods of topological quantum field theory and here we can give
only a short sketch of this methods. Therefore it seems
worthwhile to present a direct proof of this fact.
We denote the Hamiltonian of eq.~(\ref{4.1}) by $H_{N\dots 21}$ and
by $H_{1N\dots 2}$ that one obtained by the cyclic permutation
$(N\dots 21)\mapsto(1N\dots 2)$. The physical content is
invariant under this permutation since both Hamiltonians are equivalent
%//{4.3a}
\beq \label{4.3a}
R^{-1}_{(N\dots 2)1}H_{N\dots 21}R_{1(N\dots 2)}=H_{1N\dots 2}
\eeq
where, as a generalization of eq.~(\ref{2.6}), we have introduced the
R-matrices
%//{4.3b}
\beq \label{4.3b}
R_{1(N\dots 2)}\equiv
\ba{c}
%r2
\unitlength=0.50mm
%\special{em:linewidth 0.4pt}
%\linethickness{0.4pt}
\begin{picture}(50.00,30.00)
\put(20.00,5.00){\makebox(0,0)[cc]{$\s N$}}
\put(30.00,5.00){\makebox(0,0)[cc]{$\s N-1$}}
\put(50.00,5.00){\makebox(0,0)[cc]{$\s 2$}}
\put(10.00,5.00){\makebox(0,0)[cc]{$\s 1$}}
\put(20.00,10.00){\vector(-1,2){10.00}}
\put(30.00,10.00){\vector(-1,2){10.00}}
\put(50.00,10.00){\vector(-1,2){10.00}}
\put(47.50,25.00){\oval(5.00,10.00)[rb]}
\put(12.50,15.00){\oval(5.00,10.00)[lt]}
\put(50.00,25.00){\vector(0,1){5.00}}
\put(10.00,10.00){\line(0,1){5.00}}
\put(17.00,20.00){\line(1,0){6.00}}
\put(27.00,20.00){\line(1,0){2.00}}
\put(43.00,20.00){\line(-1,0){4.00}}
\put(33.00,20.00){\makebox(0,0)[cc]{$\s \cdots$}}
\end{picture}
\ea
\quad{\rm and}\quad R^{-1}_{(N\dots 2)1}\equiv
\ba{c}
%r1
\unitlength=0.50mm
%\special{em:linewidth 0.4pt}
%\linethickness{0.4pt}
\begin{picture}(50.00,33.00)
\put(10.00,10.00){\vector(1,2){10.00}}
\put(20.00,10.00){\vector(1,2){10.00}}
\put(23.00,20.00){\line(-1,0){6.00}}
\put(12.00,26.50){\oval(4.00,13.00)[lb]}
\put(10.00,25.00){\vector(0,1){5.00}}
\put(40.00,10.00){\vector(1,2){10.00}}
\put(27.00,20.00){\line(1,0){3.00}}
\put(43.00,20.00){\line(-1,0){4.00}}
\put(34.00,20.00){\makebox(0,0)[cc]{$\s \cdots$}}
\put(48.00,17.50){\oval(4.00,5.00)[rt]}
\put(50.00,18.00){\line(0,-1){8.00}}
\put(10.00,5.00){\makebox(0,0)[cc]{$\s N$}}
\put(20.00,5.00){\makebox(0,0)[cc]{$\s N-1$}}
\put(40.00,5.00){\makebox(0,0)[cc]{$\s 2$}}
\put(50.00,5.00){\makebox(0,0)[cc]{$\s 1$}}
\end{picture}
\ea\ .
\eeq
For all terms of eq.~(\ref{4.1}), except
for that one where $i=1$, the claim follows
from eq.~(\ref{2.11}) and the special Yang-Baxter relation
%//{4.3c}
\beq \label{4.3c}
R_{12}R_{13}(PP^{\Lambda_2})_{23}=(PP^{\Lambda_2})_{23}R_{13}R_{12}
\eeq
which is a consequence of the general Yang-Baxter relation (\ref{2.1})
for $x_{23}=1/q$ and $x_1\to\infty$.
For the remaining term we use the identity
%//{4.3d}
\beq \label{4.3d}
R^{-1}P^{\Lambda_2}R=RP^{\Lambda_2}R^{-1}\quad{\rm or}\quad
\ba{c}
%r4
\unitlength=0.25mm
%\linethickness{0.4pt}
\begin{picture}(80.00,70.00)
\put(10.00,20.00){\oval(20.00,20.00)[t]}
\put(10.00,50.00){\oval(20.00,20.00)[b]}
\put(10.00,30.00){\line(0,1){10.00}}
\put(20.00,60.00){\vector(0,1){10.00}}
\put(20.00,0.00){\vector(0,1){10.00}}
\put(28.00,50.00){\oval(56.00,10.00)[lt]}
\put(28.00,20.00){\oval(56.00,10.00)[lb]}
\put(25.00,55.00){\vector(1,0){8.00}}
\put(32.00,15.00){\vector(-1,0){8.00}}
\put(40.00,35.00){\makebox(0,0)[cc]{$=$}}
\put(70.00,20.00){\oval(20.00,20.00)[t]}
\put(70.00,50.00){\oval(20.00,20.00)[b]}
\put(70.00,30.00){\line(0,1){10.00}}
\put(76.00,50.00){\oval(32.00,10.00)[lt]}
\put(84.00,55.00){\vector(1,0){8.00}}
\put(76.00,20.00){\oval(32.00,10.00)[lb]}
\put(92.00,15.00){\vector(-1,0){8.00}}
\put(80.00,50.00){\vector(0,1){20.00}}
\put(80.00,0.00){\line(0,1){20.00}}
\put(80.00,0.00){\vector(0,1){8.00}}
\end{picture}
\ea
\qquad{\rm where}\quad
\ba{c}
%r5
\unitlength=0.25mm
%\linethickness{0.4pt}
\begin{picture}(20.00,60.00)
\put(10.00,10.00){\oval(20.00,20.00)[t]}
\put(10.00,40.00){\oval(20.00,20.00)[b]}
\put(10.00,20.00){\line(0,1){10.00}}
\put(0.00,40.00){\vector(0,1){10.00}}
\put(20.00,40.00){\vector(0,1){10.00}}
\put(20.00,0.00){\vector(0,1){10.00}}
\put(0.00,0.00){\vector(0,1){10.00}}
\end{picture}
\ea
=P^{\Lambda_2}.
\eeq
This relation may be obtained directly from eq.~(\ref{4.2}). It also follows
from the defining relation $\Delta_{21}R_{12}=R_{12}\Delta_{12}$
for the R-matrix of quasitriangular
Hopf algebras with the coproduct $\Delta$.
For the ($i=1$)-term in eq.~(\ref{4.1}) we obtain with eqs.~(\ref{4.3c})
and (\ref{4.3d})
%//{4.3e}
\beq \label{4.3e}
\ba{c}
%r3
\unitlength=0.50mm
%\special{em:linewidth 0.4pt}
%\linethickness{0.4pt}
\begin{picture}(140.00,78.00)
\put(20.00,5.00){\makebox(0,0)[cc]{$\s N$}}
\put(30.00,5.00){\makebox(0,0)[cc]{$\s N-1$}}
\put(50.00,5.00){\makebox(0,0)[cc]{$\s 2$}}
\put(10.00,5.00){\makebox(0,0)[cc]{$\s 1$}}
\put(20.00,10.00){\vector(-1,2){10.00}}
\put(30.00,10.00){\vector(-1,2){10.00}}
\put(50.00,10.00){\vector(-1,2){10.00}}
\put(47.50,25.00){\oval(5.00,10.00)[rb]}
\put(12.50,15.00){\oval(5.00,10.00)[lt]}
\put(50.00,25.00){\vector(0,1){5.00}}
\put(10.00,10.00){\line(0,1){5.00}}
\put(17.00,20.00){\line(1,0){6.00}}
\put(27.00,20.00){\line(1,0){2.00}}
\put(43.00,20.00){\line(-1,0){4.00}}
\put(33.00,20.00){\makebox(0,0)[cc]{$\s \cdots$}}
\put(45.00,35.00){\oval(10.00,10.00)[t]}
\put(45.00,50.00){\oval(10.00,10.00)[b]}
\put(40.00,35.00){\line(0,-1){5.00}}
\put(50.00,35.00){\line(0,-1){5.00}}
\put(45.00,40.00){\line(0,1){5.00}}
\put(40.00,50.00){\line(0,1){5.00}}
\put(50.00,50.00){\line(0,1){5.00}}
\put(20.00,30.00){\line(0,1){25.00}}
\put(10.00,30.00){\line(0,1){25.00}}
\put(10.00,55.00){\vector(1,2){10.00}}
\put(20.00,55.00){\vector(1,2){10.00}}
\put(23.00,65.00){\line(-1,0){6.00}}
\put(12.00,71.50){\oval(4.00,13.00)[lb]}
\put(10.00,70.00){\vector(0,1){5.00}}
\put(40.00,55.00){\vector(1,2){10.00}}
\put(27.00,65.00){\line(1,0){3.00}}
\put(43.00,65.00){\line(-1,0){4.00}}
\put(34.00,65.00){\makebox(0,0)[cc]{$\s \cdots$}}
\put(47.50,62.50){\oval(5.00,5.00)[rt]}
\put(50.00,55.00){\line(0,1){7.00}}
\put(70.00,40.00){\makebox(0,0)[cc]{$=$}}
\put(95.00,35.00){\oval(10.00,10.00)[t]}
\put(95.00,50.00){\oval(10.00,10.00)[b]}
\put(90.00,35.00){\line(0,-1){5.00}}
\put(100.00,35.00){\line(0,-1){5.00}}
\put(95.00,40.00){\line(0,1){5.00}}
\put(90.00,50.00){\line(0,1){5.00}}
\put(100.00,50.00){\line(0,1){5.00}}
\put(92.50,55.00){\oval(5.00,10.00)[lt]}
\put(93.50,30.00){\oval(7.00,8.00)[lb]}
\put(100.00,55.00){\vector(0,1){20.00}}
\put(100.00,10.00){\line(0,1){21.00}}
\put(110.00,10.00){\vector(0,1){65.00}}
\put(130.00,10.00){\vector(0,1){65.00}}
\put(137.50,65.00){\oval(5.00,10.00)[rb]}
\put(137.50,23.50){\oval(5.00,5.00)[rt]}
\put(98.00,26.00){\line(-1,0){4.00}}
\put(102.00,26.00){\line(1,0){6.00}}
\put(108.00,60.00){\line(-1,0){6.00}}
\put(98.00,60.00){\line(-1,0){5.00}}
\put(112.00,60.00){\line(1,0){3.00}}
\put(128.00,60.00){\line(-1,0){3.00}}
\put(128.00,26.00){\line(-1,0){3.00}}
\put(112.00,26.00){\line(1,0){3.00}}
\put(140.00,10.00){\line(0,1){14.00}}
\put(137.00,26.00){\line(-1,0){5.00}}
\put(137.00,60.00){\line(-1,0){5.00}}
\put(140.00,64.00){\vector(0,1){11.00}}
\put(119.00,26.00){\makebox(0,0)[cc]{$\s \cdots$}}
\put(119.00,60.00){\makebox(0,0)[cc]{$\s \cdots$}}
\put(140.00,5.00){\makebox(0,0)[cc]{$\s 2$}}
\put(110.00,5.00){\makebox(0,0)[cc]{$\s N$}}
\put(100.00,5.00){\makebox(0,0)[cc]{$\s 1$}}
\put(130.00,5.00){\makebox(0,0)[cc]{$\s 3$}}
\end{picture}
\ea
\eeq
which finishes the proof of eq.~(\ref{4.3a}).

We now use the results of Sect.~3, especially, the Bethe ansatz
equations (\ref{3.20}) to investigate the finite size behaviour
of the ground state energy.
In the thermodynamic limit conformal invariance of the system is
expected.
As shown by Cardy \cite{cardy} conformal invariance implies for the maximal
eigenvalue $\Lambda_{\max}$ of the transfer matrix
%//{4.4}
\beq \label{4.4}
\Lambda_{\max} \approx \exp\Big(-Nf+\frac{1}{N}\frac{\pi}{6}c\Big)
,\qquad(N \to \infty)
\eeq
where $f$ is the free energy per site and $c$ is the central charge of the
corresponding Virasoro algebra. Using the techniques developed in \cite{karfs}
the central charge can be calculated from the finite size
behaviour of $\Lambda_{\max}$ belonging to the antiferromagnetic ground state.

Taking the logarithm of the Bethe ansatz equation (\ref{3.20}) we obtain
%//{4.5}
\beq \label{4.5}
z_k(u_j^{(k)})= 2 \pi I_j^{(k)},\quad\left\{
\ba{l} I_j^{(k)} \in ({\bf Z}+\frac12)\
     \cap\ ]z_k(-\infty)/2\pi ,z_k(\infty)/2\pi [\\
 j=1,\dots,r_k,\ k=1,\dots,n-1,
\ea\right.
\eeq
where we have introduced the ``rapidities"
$u_i^{(k)}=i\gamma(k-n)+2\theta_i^{(k)}$ and the phase functions
%//{4.6}
\beq \label{4.6}
z_k(u)= N\delta_{k,n-1}p(u)+\sum_{l\in L_k}\sum_i p(u-u_i^{(l)})
+\sum_{i} \Phi(u-u_i^{(k)})+(2+\eta_k)\gamma,
\eeq
where $L_k=\{l=k\pm 1;\ 0<l<n\}$.
The inhomogeneities $x_i$ have been taken to be equal to one.
The functions $p(u)$ and $\Phi(u)$ are given by
%//{4.7}
\beq \label{4.7}
e^{ip(u)}=\frac{\sinh \frac12(u-i\gamma)}{\sinh \frac12(u+i\gamma)}
\quad {\rm and}\quad
e^{i\Phi(u)}=\frac{\sinh \frac12(u+2i\gamma)}{\sinh \frac12(u-2i\gamma)}.
\eeq
The Fourier transforms $\tilde f(x)=\int du/(2\pi) e^{iux} f(u)$
of the derivatives $p'$ and $\Phi'$ are
%//{4.8}
\beq \label{4.8}
\tilde {p'}(x)=\frac{\sinh(\pi-\gamma)x}{\sinh \pi x}
\quad {\rm and}\quad
\tilde {\Phi'}(x)=1-2 \cosh\gamma x \ \tilde {p'}(x).
\eeq
The eqs.~(\ref{4.6}-\ref{4.8}) may be generalized to include
string solutions as well. However,
since we want to analyse the finite size behaviour of the ground state
eigen value, we are only looking for real roots of the Bethe ansatz equations
(\ref{4.5}).
Taking the derivative of eq.~(\ref{4.6}) we find the following matrix
equation
%//{4.8a}
\beq \label{4.8a}
z'_k=\rho_k+\varphi_k=N\delta_{k,n-1}p'+\sum_l(p'+\Phi')_{kl}*\rho_l
\eeq
where the matrices $p'$ and $\Phi'$ are given
by $p'_{kl}=\sum_{l'\in L_k}\delta_{l'l}\,
p'$ and $\Phi'_{kl}=\delta_{kl}\Phi'$, respectively.
The convolution is defined by $(f*g)(u)=\int du'/(2\pi)f(u-u')g(u')$.
The densities of roots
$\rho_k(u)=2\pi\sum_{i=1}^{r_k}\delta(u-u_i^{(k)})$ may be written in
terms of
$\varphi_k(u)$, which describes the density of Bethe ansatz holes
as well as finite size corrections. For this purpose one has to
invert the matrix $(1-p'-\Phi')_{kl}$
appearing in the integral equation (\ref{4.8a}).
Its Fourier transform is given by the symmetric matrix
%//{4.9}
\beq \label{4.9}
\widetilde{(1-p'-\Phi')}_{kl}^{-1}(x)=\frac{\sinh\pi x\ \sinh(n-k)\gamma x
\ \sinh l\gamma x}{\sinh(\pi-\gamma)x\ \sinh n\gamma x\ \sinh\gamma x},
\quad k\le l.
\eeq

For large lattice size $N$ and $x\approx 1$ the eigenvalue $\Lambda(x)$
(see eq.~(\ref{3.18})) of the transfer matrix
is dominated by the term $\lambda_n(x)$.
The following calculation may be easily performed also for
excitations but for simplicity we restrict here to the ground state.
{}From eqs.~(\ref{3.19}), (\ref{4.7}) and (\ref{4.8a}) we obtain
%//{4.10}
\begin{eqnarray} \label{4.10}
\lefteqn{\log\lambda_n(\theta,\gamma)
 =N\log a-\int\frac{du}{2\pi}\ ip(u-2\theta)\ \rho_{n-1}(u)-i\gamma}\nonumber\\
&&=-Nf_{\infty}+\int\frac{du}{2\pi}\int\frac{du'}{2\pi}\sum_{k=1}^{n-1}
 ip(u-2\theta)\,\big(1-p'-\Phi'\big)^{-1}_{n-1,k}(u-u')\,
\varphi_k(u)-i\gamma,\quad
\end{eqnarray}
where
$f_\infty$ is the free energy per site in the thermodynamic limit.
Using the techniques of ref.~\cite{karfs} we find with eq.~(\ref{4.4})
for the central charge of the Virasoro algebra the formula
%//{4.12}
\beq \label{4.12}
c=\sum_{k,l=1}^{n-1}\bigg(\delta_{kl}-\frac{12}{r^2}\ \widetilde{
(1-p'-\Phi')}^{-1}_{kl}(0)\bigg).
\eeq
Taking eq.~(\ref{4.9}) at $x=0$ we can perform the sums and obtain
%//{4.13}
\beq \label{4.13}
c=(n-1)\left(1-\frac{n(n+1)}{r(r-1)}\right)
, \qquad q=e^{i\pi/r},\quad r=n+2,n+3,\ldots.
\eeq
This formula has previously been obtained in ref.~\cite{japan} by means of
Baxters \cite{baxter} corner transfer matrix method for the $A_{n-1}$
RSOS-models.
Note that the matrix $\widetilde{(1-p'-\Phi')}(0)$
is just $\frac{r}{r-1}$ times
the Cartan matrix $A$ (see e.g.~\cite{wieg}).
We expect the central charge of models
for general simply laced $q$-Lie algebras ($A,D,E$) of rank $l$ to be
%//{4.14}
\beq \label{4.14}
c=\sum_{i,j=1}^{l}\bigg(\delta_{ij}-\frac{12}{r(r-1)}
A^{-1}_{ij}\bigg).
\eeq
One easily can calculate the sums and finds
%//{4.15}
\beq \label{4.15}
c=l\left(1-\frac{g(g+1)}{r(r-1)}\right)
, \qquad q=e^{i\pi/r},\quad r=g+2,g+3,\ldots,
\eeq
where $g$ is the dual Coxeter number
(for $A_l,D_l,E_6,E_7,E_8$: $g=l+1,2l-2,12,18,30$, respectively).
This formula coincides (see \cite{bouwknegt}) with the formula for
the central charges of the extended coset algebras constructions
of ref.~\cite{gko} for general simply laced Lie algebras.

In a forthcoming paper we will in addition to the ground
state also discuss the excitation spectrum, also for other models related to
other quantum groups. As mentioned in Appendix A the quantum group
representation of the states are determined by the topology of the interior
of the 3-manifold, on whose boundary the vertex model is defined.
For the case of trivial topology of the interior of a torus there
exist only states which transform
trivially under the quantum group. We will analyse more general situations
to obtain higher representations.
In this paper we have not mentioned questions
about positivity, unitarity e.t.c., these will be investigated elsewhere.
\section*{Appendix A}
In this appendix we define
the vertex model  with periodic boundary conditions of eq.~(\ref{1.0})
depicted in Fig.~\ref{f1}.
The transfer matrix of eq.~(\ref{3.3}) is shown to belong
to this model. We use
the techniques of topological quantum field theory in terms of
coloured graphs as developed in ref. \cite{ks}.
Here we restrict the derivation to the quantum
group $Sl_q(2)$ for $q=\exp (i\pi/r),\ (r=3,4,5,\dots)$.
The construction may easily be generalized to other quantum groups
(see e.g. \cite{belia}). The model is only defined for $q$ equal
to roots of unity, whereas the transfer matrix (\ref{3.3})
is of course also meaningful for generic values of $q$.

Let $M$ be a 3-manifold, a solid torus or more general a cylindric part of a
handle of an arbitrary handle body.
On the boundary $\partial M$ of $M$
we consider a graph $G(\underline x)$ $(\underline x=\{x_1,\dots,x_N,x\}$)
which forms a square lattice as in Fig.~\ref{f1} or on the left hand side of
Fig.~\ref{fa}.
\begin{figure}%[h]
$$
\begin{array}{c}
\setlength{\unitlength}{.8mm}
\begin{picture}(90.,70.)
%\input fa1
%fa1
\put(5.,60.25){\line(0,-1){45.}}
\put(50.,15.25){\line(0,1){45.}}
\bezier{200}(5.,15.25)(27.50,0.25)(50.,15.25)
\bezier{200}(5.,60.25)(27.50,45.25)(50.,60.25)
\bezier{200}(5.,60.25)(27.50,75.25)(50.,60.25)
\thicklines
\put(70.,37.5){\vector(1,0){15.}}
\bezier{30}(5.,30.25)(27.50,15.25)(49.50,30.25)
\bezier{30}(5.,44.75)(27.50,29.75)(49.50,44.75)
\bezier{100}(5.,23.)(27.50,8.)(49.50,23.)
\bezier{100}(5.,37.50)(27.50,22.50)(49.50,37.50)
\bezier{100}(5.,53.)(27.50,38.)(49.50,53.)
\put(51.,22.){\makebox(6.25,3.75){$x$}}
\put(51.,36.5){\makebox(6.25,3.75){$x$}}
\put(51.,52.){\makebox(6.25,3.75){$x$}}
\put(10.,57.20){\line(0,-1){45.}}
\put(20.,53.60){\line(0,-1){45.}}
\put(35.,53.60){\line(0,-1){45.}}
\put(45.,57.20){\line(0,-1){45.}}
\thinlines
\put(6.50,2.25){\makebox(6.50,6.){$x_2$}}
\put(16.75,1.25){\makebox(6.25,3.75){$x_1$}}
\put(32.50,1.75){\makebox(5.25,3.){$x_N$}}
\put(50.50,10.75){\makebox(2.,1.75){$\cdot$}}
\put(3.75,7.25){\makebox(2.,1.75){$\cdot$}}
\put(0.75,10.25){\makebox(2.,1.75){$\cdot$}}
\put(41.75,4.50){\makebox(2.,1.75){$\cdot$}}
\put(46.50,7.25){\makebox(2.,1.75){$\cdot$}}
\put(59.,45.){\makebox(7.75,6.25){$G(\underline x)$}}
\put(56.,48.){\vector(-4,1){9.}}
\put(-1.,33.){\makebox(1.,1.){$M$}}
\end{picture}
\end{array}
\begin{array}{c}
\setlength{\unitlength}{.8mm}
\begin{picture}(80.,100.)
%\input fa2
%fa2.tex
\put(-2.,76.){\makebox(7.75,6.25){$M^{(1)}$}}
\put(64.50,72.75){\makebox(7.75,6.25){$G^{(1)}(\underline x)$}}
\put(58.,77.){\vector(-4,1){9.}}
\put(8.,90.){\line(0,-1){15.}}
\put(53.,75.25){\line(0,1){14.75}}
\bezier{200}(8.,90.)(28.,105.)(53.,90.)
\bezier{200}(8.,90.)(28.,75.)(53.,90.)
\bezier{200}(7.75,75.)(27.75,60.)(52.75,75.)
\thicklines
\bezier{150}(8.,82.75)(28.,67.25)(53.,82.50)
\put(13.,86.75){\line(0,-1){15.25}}
\put(23.,83.){\line(0,-1){15.25}}
\put(38.,83.25){\line(0,-1){14.75}}
\put(48.,87.25){\line(0,-1){15.}}
\thinlines
\put(-2.,43.){\makebox(7.75,6.25){$M^{(2)}$}}
\put(8.,57.75){\line(0,-1){15.}}
\put(53.,43.){\line(0,1){14.75}}
\put(19.25,54.25){\makebox(7.25,5.50){$\scriptstyle a'_1$}}
\put(26.50,58.2){\makebox(6.25,5.25){$\cdots$}}
\put(30.,54.50){\makebox(8.25,5.75){$\scriptstyle a'_{N-1}$}}
\put(64.25,39.50){\makebox(7.75,6.25){$G^{(2)}(\underline x)$}}
\put(58.,44.){\vector(-4,1){9.}}
\put(56.,61.75){\makebox(10.50,7.50){$G^{D^2}_{\underline a'}$}}
\put(54.50,63.50){\vector(-3,-1){10.25}}
\bezier{200}(8.,57.75)(28.,72.75)(53.,57.75)
\bezier{200}(8.,57.75)(28.,42.75)(53.,57.75)
\bezier{200}(7.75,42.75)(27.75,27.75)(52.75,42.75)
\thicklines
\bezier{150}(24.50,53.75)(6.,61.)(30.50,62.50)
\bezier{150}(30.50,62.50)(53.50,59.75)(35.75,54.25)
\bezier{20}(23.,50.75)(23.75,52.25)(24.50,53.75)
\bezier{20}(38.,51.25)(36.62,52.88)(35.75,54.25)
\put(48.,55.){\line(-4,1){6.50}}
\put(13.25,54.50){\line(3,1){5.75}}
\bezier{150}(8.,50.50)(28.,35.)(53.,50.25)
\put(13.,54.50){\line(0,-1){15.25}}
\put(23.,50.75){\line(0,-1){15.25}}
\put(38.,51.){\line(0,-1){14.75}}
\put(48.,55.){\line(0,-1){15.}}
\thinlines
\put(-2.,10.){\makebox(7.75,6.25){$M^{(3)}$}}
\put(64.,6.){\makebox(7.75,6.25){$G^{(3)}(\underline x)$}}
\put(58.,11.){\vector(-4,1){9.}}
\put(8.,24.75){\line(0,-1){15.}}
\put(53.,10.){\line(0,1){14.75}}
\bezier{200}(8.,24.75)(28.,39.75)(53.,24.75)
\bezier{200}(8.,24.75)(28.,9.75)(53.,24.75)
\bezier{200}(7.75,9.75)(27.75,-5.25)(52.75,9.75)
\put(19.50,21.25){\makebox(6.25,5.25){$\scriptstyle a_1$}}
\put(26.50,25.2){\makebox(6.25,5.25){$\cdots$}}
\put(30.,21.20){\makebox(7.25,5.){$\scriptstyle a_{N-1}$}}
\put(56.,28.50){\makebox(10.50,7.50){$G^{D^2}_{\underline a}$}}
\put(54.50,31.25){\vector(-3,-1){10.75}}
\thicklines
\bezier{200}(8.,17.50)(28.,2.)(53.,17.25)
\put(13.,21.50){\line(0,-1){15.25}}
\put(23.,17.75){\line(0,-1){15.25}}
\put(38.,18.){\line(0,-1){14.75}}
\put(48.,22.){\line(0,-1){15.}}
\bezier{200}(24.50,20.75)(6.,28.)(30.50,29.50)
\bezier{200}(30.50,29.50)(53.50,26.75)(35.75,21.25)
\bezier{50}(23.,17.75)(23.75,19.25)(24.50,20.75)
\bezier{50}(38.,18.25)(36.62,19.88)(35.75,21.25)
\put(48.,22.){\line(-4,1){6.50}}
\put(13.25,21.50){\line(3,1){5.75}}
\end{picture}
\end{array}
$$
\caption{\label{fa}\it Surgery along the dotted lines}
\end{figure}
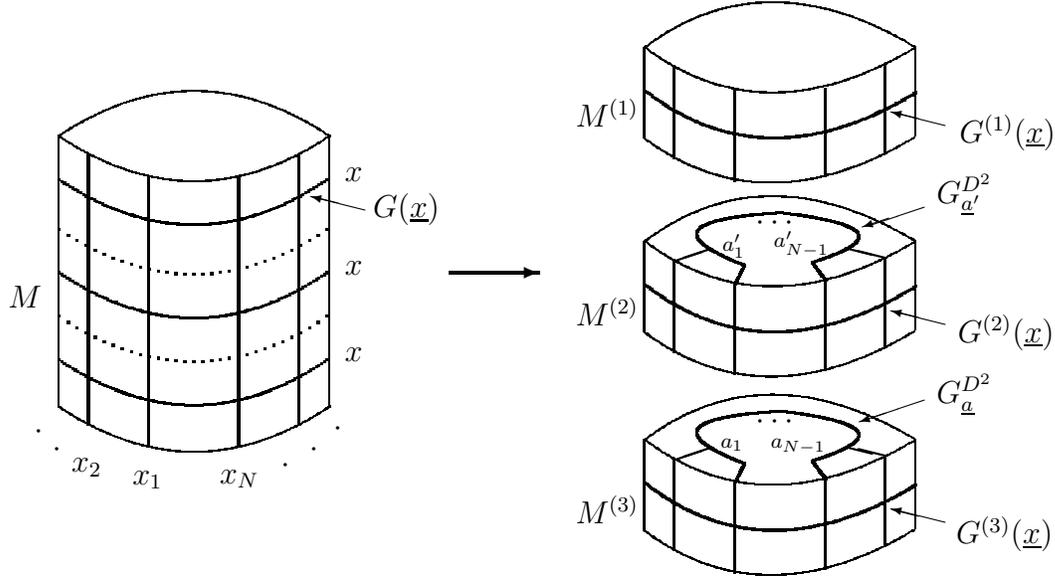
To each line of the graph we associate the fundamental representation
$\Lambda_1$ and a spectral parameter
$x_i$ or $x$ which coincide on opposite legs of the 4-vertices.
The 4-vertices are given by eq.~(\ref{2.2}) in terms of spectral parameter
independent R-matrices. For simplicity we take the spectral parameters
of the horizontal lines all equal to $x$.
Formula (3.2) of ref.~\cite{ks} defines a partition function
%
%//{a.1}
\begin{equation}\label{a.1}
Z(M,G(\underline x))=\sum_{\underline j,\underline{\tilde J}}
W(\underline j,\underline{\tilde J})(X,G(\underline x)).
\end{equation}
The right hand side is defined in terms of a
triangulation $X$ of the 3-manifold $M$ inducing a triangulation
$\partial X$ of $\partial M$.
The sum of eq.~(\ref{a.1}) runs over all set of colours $\underline j$ and
$\underline{\tilde J}$, where the colours are the irreducible
representations ($j=0,1/2,1,\dots,r/2-1$) of $Sl_q(2)$.
The set $\underline j$ is a colouring
of all 1-simplexes of $X$ and $\underline{\tilde J}$ a colouring of
all plaquettes obtained from the graph $\widetilde{\partial X}\cup
G(\underline x)$, where $\widetilde{\partial X}$ is the dual
graph of $\partial X$. In ref.~\cite{ks} it is explained how the weights
$W(\underline j,\underline{\tilde J})(X,G(\underline x))$ are given
in terms of q-dimensions, 6j-symbols and R-matrices.
Moreover it is shown that the r.h.s. of eq.~(\ref{a.1}) does not
depend on the triangulation $X$ of $M$ and therefore defines an invariant
of the 3-manifold M equipped with a vertex model on its boundary
$\partial M$ given by the graph $G(\underline x)$.

Note that the partition function (\ref{a.1}) does not only describes
a two dimensional vertex model on the torus or cylinder but in
addition there is a local interaction of the vertices with the interior
of the 3-manifold $M$. However, this interaction is of topological
nature. Thus this model is similar to $\sigma$-models with Chern-Simons
term or the WZNW-models \cite{nov}\cite{wit}.

We decompose $M$ and the square lattice represented by the
graph  $G(\underline x)$ on the boundary as follows
%
%//{a.2}
\begin{eqnarray}\label{a.2}
M&=&M^{(1)}\cup_{D^2}M^{(2)}\cup_{D^2}M^{(3)}\nonumber \\
G(\underline x)&=&G^{(1)}(\underline x)\cup G^{(2)}(\underline x)
 \cup G^{(3)}(\underline x)
\end{eqnarray}
along two discs $D^2$ as shown in Fig.~\ref{fa}.
Applying the general surgery formula (7.3) of ref. \cite{ks} we have
%//{a.3}
\begin{eqnarray} \label{a.3}
 Z(M,G(\underline x))&=& \sum_{\underline a\underline a'}
 W^{D^2}_{\underline a} W^{D^2}_{\underline a'}\
 Z(M^{(1)},G^{(1)}(\underline x)\cup G^{D^{2*}}_{\underline a'})\
\nonumber \\&&\qquad \times
 Z(M^{(2)},G^{(2)}(\underline x)\cup G^{D^{2}}_{\underline a'}\cup
  G^{D^{2*}}_{\underline a})\
 Z(M^{(3)},G^{(3)}(\underline x)\cup G^{D^{2}}_{\underline a}) .
\end{eqnarray}
where $G^{D^2}_{\underline a}$ as depicted in Fig.~\ref{fa},
the canonical graph of the disc $D^2$, is defined in ref. \cite{ks}.
On the bottoms of $M^{(1)}$ and $M^{(2)}$ there are the
mirror graphs $G^{D^{2*}}_{\underline a'}$ and  $G^{D^{2*}}_{\underline a}$,
respectively. The (finite) summation is over all colourings
$\underline a$ and $\underline a'$ of the canonical graphs.
Note that the following also holds, if $M^{(1)}$ and $M^{(3)}$ stay
connected after the surgery.
The piece $M^{(2)}$ is topological equivalent to the ball
$D^3$ with boundary $\partial D^3=S^2$ and the graph
$G^{(2)}(\underline x)\cup G^{D^{2}}_{\underline a'}\cup
  G^{D^{2*}}_{\underline a}$ is planar. In ref. \cite{ks} is shown that
for planar graphs the interaction with the interior of the 3-manifold
disappears and
%//{a.4}
%\vskip2mm
\begin{equation}\label{a.4}
 Z(M^{(2)},G^{(2)}(\underline x)\cup G^{D^{2}}_{\underline a'}\cup
  G^{D^{2*}}_{\underline a})=
\tau^{\underline a'}_{\underline a}(x,\{x_i\})\equiv
\ba{c}
\unitlength=0.50mm
\begin{picture}(100.,50.)
\put(55.,15.){\line(1,0){25.}}
\put(40.,15.){\makebox(0,0)[cc]{$\cdots$}}
\put(25.,5.){\makebox(0,0)[cc]{$\s x_N$}}
\put(65.,5.){\makebox(0,0)[cc]{$\s x_2$}}
\put(75.,5.){\makebox(0,0)[cc]{$\s x_1$}}
\put(85.,15.){\makebox(0,0)[lb]{$\s x$}}
%\put(20.,15.){\oval(10.,10.)[rb]}
%\put(60.,15.){\oval(10.,10.)[rb]}
%\put(70.,15.){\oval(10.,10.)[rb]}
\put(15.,25.){\oval(20.,20.)[l]}
\put(20.,0.){\line(1,0){50.}}
\put(20.,45.){\line(1,0){50.}}
\put(45.,-5.){\makebox(0,0)[cc]{$\s \underline a$}}
\put(45.,51.){\makebox(0,0)[cc]{$\s \underline a'$}}
\put(70.,0.){\line(0,1){45.}}
\put(60.,0.){\line(0,1){45.}}
\put(20.,0.){\line(0,1){45.}}
\put(50.,35.){\line(1,0){5.}}
\put(50.,15.){\line(1,0){5.}}
\put(40.,35.){\makebox(0,0)[cc]{$\cdots$}}
\put(5.,25.){\makebox(0,0)[cc]{-}}
\put(22.,35.){\line(1,0){8.}}
\put(50.,35.){\line(1,0){8.}}
\put(62.,35.){\line(1,0){6.}}
\put(72.,35.){\line(1,0){8.}}
\put(80.,35.){\line(0,0){0.}}
\put(80.,40.){\oval(10.,10.)[rb]}
\put(80.,10.){\oval(10.,10.)[rt]}
\put(92.50,40.){\oval(15.,20.)[t]}
\put(92.50,10.){\oval(15.,20.)[b]}
\put(100.,25.){\makebox(0,0)[cc]{-}}
\put(30.,15.){\line(-1,0){16.}}
\put(14.,35.){\line(1,0){4.}}
\put(74.,35.){\line(1,0){4.}}
\put(100.,41.){\line(0,-1){12.}}
\put(100.,9.){\line(0,1){22.}}
\end{picture}
\ea,
\end{equation}
\vskip2mm\noindent
where the 3-vertices are given by intertwiners $V^{a_i}\otimes V^{1/2}\to
V^{a_{i+1}}$ as explained in ref. \cite{ks}. The transfer matrix
$\tau^{\underline a'}_{\underline a}(x,\{x_i\})$ given by eq.~(\ref{a.4})
 is represented in the path basis. It is equivalent to the transfer matrix
$\tau^{\underline \alpha'}_{\underline \alpha}(x,\{x_i\})$ given by
eq.~(\ref{3.3}),
represented in the tensor basis, projected to the sector of total spin
$J=0$. We remark that the other sectors are obtained for nontrivial
topology of the interior of the 3-manifold $M$. In a forthcoming
paper we will discuss this more general situation. We stress again
that the invariant
$Z(M^{(2)},G^{(2)}(\underline x)\cup G^{D^{2}}_{\underline a'}
\cup G^{D^{2*}}_{\underline a})$ of
eq.~(\ref{a.4}) is defined only for $q$ equal to roots of unity,
whereas the transfer matrices
$\tau^{\underline a'}_{\underline a}(x,\{x_i\})$ and
$\tau^{\underline \alpha'}_{\underline \alpha}(x,\{x_i\})$ are also
meaningful for generic values of $q$.
\section*{Appendix B}
We define the matrices $L^\pm$ by the limits $x$ to $\infty$ or $0$ of the
monodromy matrix $T$ given by eq.~(\ref{3.1})
%//{b.1a}
\beq \label {b.1a}
L^+=\lim_{x \to \infty} x^{-N} T(x)
= \left( \ba{cccc} 1 & 0 & \cdots & 0\\
\alpha E_1 &1 &\ddots &\vdots \\ & \ddots & \ddots & 0 \\
\ast & & \alpha E_{n-1}& 1 \ea \right)
\left( \ba{cccc} q^{W_1}&0&  \cdots & 0\\ 0& q^{W_2} &\ddots&\vdots\\
\vdots & \ddots & \ddots & 0\\ 0 & \cdots &0 & q^{W_n} \ea \right)
\eeq
%//{b.1b}
\beq \label {b.1b}
L^-=\lim_{x \to 0}x^N T(x)
= \left( \ba{cccc} q^{-W_1}&0&  \cdots & 0\\ 0& q^{-W_2} &\ddots&\vdots\\
\vdots & \ddots & \ddots & 0\\ 0 & \cdots &0 & q^{-W_n} \ea \right)
\left( \ba{cccc} 1 & -\alpha F_1 & & \ast \\
0 &1 &\ddots & \\ & \ddots & \ddots & -\alpha F_{n-1}\\
0 & & 0& 1 \ea \right).
\eeq
These forms of the matrices $L^+$ and $L^-$ follow from the triangular
form of the R-matrix in eq.~(\ref{2.3}) (see also ref.~\cite{ding}).
The entries $E_i,F_i$ and $q^{\pm W_i}$ of $L^\pm$ are given by the
N-fold coproduct of the generating elements $q^{\pm h_i/2}$,
$e_i$ and $f_i, \ (i=1,\ldots,n-1)$ in the fundamental representation
$\Lambda_1$ of $\hat{U}=U_q(sl(n))$ associated to each lattice site:
\begin{eqnarray}
E_i &=& \Delta^{(N)}(e_i)=\sum_{l=1}^N q^{-h_i} \ot \ldots \ot q^{-h_i}\ot
\underbrace{e_i}_{l \rm th}
\ot \einsop \ot \ldots \ot \einsop, \nonumber \\
F_i &=& \Delta^{(N)}(f_i)=\sum_{l=1}^N \einsop \ot \ldots \ot \einsop \ot
\underbrace{f_i}_{l\rm th}\ot q^{h_i} \ot \ldots \ot q^{h_i},\nonumber \\
q^{\pm H_i/2}&=&\Delta^{(N)} (h_i)= q^{\pm h_i/2} \ot \ldots \ot q^{\pm h_i/2},
\end{eqnarray}
The $H_i=W_i-W_{i+1}$ are the Cartan elements.
The Yang-Baxter equation for monodromy matrices implies the commutation
rules
%
%//{b.2}
\beq \label {b.2}
R L^\pm_1 L^\pm_2= L^\pm_2 L^\pm_1 R \quad \mbox{ and } \quad
R L^+_1 L^-_2 =L^-_2 L^+_1 R .
\eeq
These commutation rules are equivalent \cite{FRT} \cite{ding}
to the defining relations of $U_q(sl(n))$
for the elements $E_i,F_i$ and $q^{\pm H_i/2}$.
$\Delta^{(N)}$ is an algebra homomorphism $\hat{U} \to \hat{U} \ot
\ldots \ot \hat{U}$.
Quantum group invariance of the transfer matrix is now shown by applying
$L^\pm$, e.g.
\beq
L^+ \tau(x)=\lim_{y \to \infty} \left( \prod_{i=1}^N \frac{-y}{x_i} \right)
T(y)\ \tau(x).
\eeq
Using the Yang-Baxter (\ref{2.1}), crossing (\ref{2.10}) and inversion
(\ref{2.1}) relations one obtains
\beq
L^+\tau(x)=\ba{c}
\unitlength=0.50mm
\begin{picture}(64.,45.)
\put(20.,17.50){\oval(40.,15.)[l]}
\put(44.,17.50){\oval(40.,15.)[r]}
\put(59.,5.){\makebox(0,0)[cc]{$\s x$}}
\put(22.,0.){\vector(0,1){45.}}
\put(27.,0.){\vector(0,1){45.}}
\put(42.,0.){\vector(0,1){45.}}
\put(34.,4.){\makebox(0,0)[cc]{$\s \cdots$}}
\put(40.,35.){\line(-1,0){11.}}
\put(25.,35.){\line(-1,0){1.}}
\put(24.,25.){\line(1,0){1.}}
\put(29.,25.){\line(1,0){11.}}
\put(64.,35.){\line(-1,0){20.}}
\put(21.,35.){\vector(-1,0){21.}}
\put(50.,10.){\vector(-1,0){40.}}
\end{picture}
\ea =
%\ba{c}
%\input fa-2
%\ea =
\ba{c}
\unitlength=0.50mm
\begin{picture}(64.,45.)
\put(20.,27.50){\oval(40.,15.)[l]}
\put(44.,27.50){\oval(40.,15.)[r]}
\put(59.,15.){\makebox(0,0)[cc]{$\s x$}}
\put(22.,0.){\vector(0,1){45.}}
\put(27.,0.){\vector(0,1){45.}}
\put(42.,0.){\vector(0,1){45.}}
\put(34.,4.){\makebox(0,0)[cc]{$\s \cdots$}}
\put(40.,10.){\line(-1,0){11.}}
\put(25.,10.){\line(-1,0){1.}}
\put(24.,35.){\line(1,0){1.}}
\put(29.,35.){\line(1,0){11.}}
\put(64.,10.){\line(-1,0){20.}}
\put(21.,10.){\vector(-1,0){21.}}
\put(50.,20.){\vector(-1,0){40.}}
\end{picture}
\ea=\tau(x)L^+.
\eeq
%
%
%%%%%%%%%%%%%%%%%%%%%%%%%%%%%%%%%%%%%%%%%%%%%%%%%%%%%%%%%%%%%%%%%%%%%%%%%
%

%
\end{document}